\documentclass[useAMS,usenatbib,manuscript]{mn2e}
\usepackage{graphicx}
\usepackage{amsmath} 
\usepackage{amssymb} 
\usepackage{natbib}

\newcommand{\Eq}[1]{Eq.\,(\ref{#1})}
\newcommand{\Eqs}[1]{Eqs.~(\ref{#1})}
\newcommand{\Fig}[1]{Fig.\,\ref{#1}}
\newcommand{\Sec}[1]{Sec.\,\ref{#1}}
\newcommand{\Tab}[1]{Tab.\,\ref{#1}}

\newcommand{\ie}{i.\,e.}
\newcommand{\eg}{e.\,g.}

\newcommand{\MNUniverse}{{\it MareNostrum Universe}}

\newcommand{\hMpc}{{\ifmmode{h^{-1}{\rm Mpc}}\else{$h^{-1}$Mpc}\fi}}
\newcommand{\hkpc}{{\ifmmode{h^{-1}{\rm kpc}}\else{$h^{-1}$kpc}\fi}}
\newcommand{\hMsun}{{\ifmmode{h^{-1}{\rm {M_{\odot}}}}\else{$h^{-1}{\rm{M_{\odot}}}$}\fi}}
\newcommand{\Msun}{{\ifmmode{{\rm {M_{\odot}}}}\else{${\rm{M_{\odot}}}$}\fi}}

\def\muGauss{\ifmmode{\:\mu{\rm G}}\else{\,$\mu$G}\fi}

\renewcommand{\vec}[1]{{ \bf #1 }}

\def\up{_{\rm u}}
\def\down{_{\rm d}}

\title[Diffuse radio emission in cosmological simulations]
      {Diffuse radio emission from clusters \\ in the \MNUniverse\ simulation}

\author[Hoeft et al.]
   {
  M. Hoeft$^1$, M. Br\"uggen$^2$, G. Yepes$^3$, S. Gottl\"ober$^1$, and A. Schwope$^1$ \\
  $^1$Astrophysikalisches Institut Potsdam, An der Sternwarte 16, 14482 Potsdam, Germany\\
  $^2$Jacobs University Bremen, Campus Ring 1, 28759 Bremen, Germany \\
  $^3$Grupo de Astrof\'{\i}sica, Universidad Aut{\'o}noma de Madrid, Cantoblanco, 28039 Madrid, Spain 
  }
  
\date{}

\begin{document}

\maketitle

\begin{abstract}

   Large-scale diffuse radio emission is observed in some clusters of
   galaxies. There is ample of evidence that the emission has its origin
   in synchrotron losses of relativistic electrons, accelerated in the
   course of clusters mergers. In a cosmological simulation we locate
   the structure formation shocks and estimate their radio emission. We
   proceed as follows: Introducing a novel approach to identify strong
   shock fronts in an SPH simulation, we determine the Mach number as
   well as the downstream density and temperature in the \MNUniverse\
   simulation which has $2 \times 1024^3$ particles in a
   $500\:h^{-1}\,{\rm Mpc}$ box and was carried out with non-radiative
   physics. Then, we estimate the radio emission using the formalism
   derived in \citet{hoeft:07} and produce artificial radio maps of
   massive clusters. Several of our clusters show radio objects with
   similar morphology to large-scale radio relics found in the sky,
   whereas about half of the clusters show only very little radio
   emission. In agreement with observational findings, the maximum
   diffuse radio emission of our clusters depends strongly on their
   X-ray temperature. We find that the so-called accretion shocks cause
   only very little radio emission. We conclude that a moderate
   efficiency of shock acceleration, namely $\xi_{\rm e} \lesssim
   0.005$, and moderate magnetic fields in the region of the relics,
   namely 0.07 to $0.8\,{\rm \mu G}$ are sufficient to reproduce the
   number density and luminosity of radio relics.

\end{abstract}


\section{Introduction}
\label{sec-intro}

   The large-scale structure of the Universe, composed of clusters,
   superclusters, filaments, and sheets of galaxies, is still in the
   process of formation. Overdense regions such as clusters and
   filaments keep accreting matter. Gas streams out of cosmic voids onto
   the sheets and filaments. When the newly accreted gas collides with
   the denser gas within these structures, shock fronts arise,
   dissipating the kinetic energy. In sheets and filaments, the gas
   follows the gravitational potential towards the clusters of galaxies.
   Eventually, the gas collides with the intra-cluster medium (ICM) with
   a few $1000\:{\rm km\,s^{-1}}$ an gets heated to temperatures of
   $10^{7}$ to $10^{8}\,{\rm K}$. 
   \\

   The flow of gas is not as steady as the above picture suggests. A
   significant fraction of the gas accretion onto clusters is in the form
   of groups and clusters. The mergers of rich clusters are -- to our
   knowledge -- the most energetic events after the Big Bang. Kinetic
   energies of the order of $10^{64}$ erg are dissipated in giant shock
   waves. Only in recent years, X-ray telescopes have reached the
   necessary spatial and spectral resolution to detect the signatures of
   such shock waves in a few massive clusters. 
   \\

   A number of diffuse, steep-spectrum radio sources without optical
   identification have been observed in galaxy clusters. These sources
   have complex morphologies and show diffuse and irregular emission
   \citep{kempner:01, slee:01, bacchi:03, feretti:05, giovannini:06}. They
   are usually subdivided into two classes, denoted as `radio halos' and
   `radio relics'. Cluster radio halos are unpolarised and have diffuse
   morphologies that are similar to those of the thermal X-ray emission
   of the cluster gas \citep{giovannini:06}. Examples for clusters with
   radio halos are the Coma cluster \citep{kim:89,deiss:97}, the galaxy
   cluster 1E\,0657-56 \citep{liang:00}, the X-ray luminous cluster
   A2163, and distant clusters such as CL\,0016+16. The cluster
   A520 shows a halo with a low surface brightness with a clumpy
   structure \citep{govoni:01}. Other examples can be found in
   \citet{giovannini:99}. In general, radio halos are found in clusters
   with significant substructure and rich clusters with high X-ray
   luminosities and temperatures. The radio power correlates strongly
   with the cluster luminosity \citep{feretti:05}.
   \\

   Unlike halos, radio relics are typically located near the periphery
   of the cluster. They often exhibit sharp emission edges and many of
   them show strong radio polarisation \citep{giovannini:04}. The sizes
   of relics and the distance to the cluster centre vary significantly.
   Examples for radio relic with sizes of one Mpc or even larger have
   been observed in Coma and A2256, which contain both a relic and a
   halo (as do A225, A1300, A2744 and A754). The cluster A3667
   \citep{roettgering:97} contains two very luminous, almost symmetric
   relics with a separation of more than five Mpc. The cluster A3376
   shows an almost ring-like radio emission \citep{bagchi:06}. The
   clusters A115 and A1664 show relics only at one side of the elongated
   X-ray distribution \citep{govoni:01}. The relic source 0917+75 is
   particularly puzzling as it is located at 5 to $8\,{\rm Mpc}$ from
   the most nearby clusters. Other clusters show rather small relics as
   for example A85 \citep{slee:01}.
   \\

   The spectra of the diffuse radio sources indicate that their origin
   lies in synchrotron losses of relativistic electrons. The cooling
   time of the electrons which cause observable emission is of the order
   of one hundred Myr \citep[see \eg\ the models in][]{slee:01}. The
   origin of the relativistic electron population which causes the
   emission is still not clear. There are essentially two classes of
   models that explain the presence of relativistic electrons. Either
   they are injected into the ICM via AGN activity or stellar feedback
   or they obtain their energy from particle acceleration at large-scale
   shock fronts in galaxy clusters. As discussed above, structure
   formation in the universe causes a variety of shock fronts in the
   intergalactic medium and the ICM \citep{miniati:00, ryu:03, ryu:08}.
   These shock fronts are expected to be collisionless and capable of
   accelerating protons and electrons to relativistic energies. The
   correlation between the presence of diffuse radio emission in galaxy
   clusters and signs for a recent merger supports the scenario in which
   merger shocks generate the necessary relativistic electrons
   \citep{feretti:06}. A3367 may serve as another piece of evidence: The
   radio relic is located where the merger induced bow shock is expected
   \citep{roettiger:99}. Three mechanisms have been proposed for the
   action of the shock wave: (i) in-situ diffusive shock acceleration of
   electrons by the Fermi\,I process \citep[primary
   electrons,][]{ensslin:98, roettiger:99, miniati:01}, (ii)
   re-acceleration of electrons by compression of existing cocoons of
   radio plasma \citep{ensslin:01, ensslin:02, hoeft:04}, or (iii)
   in-situ acceleration of protons and the production of relativistic
   electrons and positrons by ineleastic p-p collisions (secondary
   electrons). In the first two cases, the diffuse radio emission is
   roughly confined to the region of the shock fronts. In contrast, in
   the latter scenario the relativistic protons have long life times and
   can travel a large distance from their source before they release
   their energy. Hence, secondary electrons may be lead to radio halos.
   \\

   In principle, observations of non-thermal cluster phenomena could
   provide an independent and complementary way of studying the growth
   of structure in our Universe and could shed light on the existence
   and the properties of the warm-hot intergalactic medium (WHIM),
   provided the underlying processes are understood. Sheets and
   filaments are predicted to host this WHIM with temperatures in the
   range $10^5$ to $10^7\:{\rm K}$ whose evolution is primarily driven
   by shock heating from gravitational perturbations breaking on mildly
   non-linear, non-equilibrium structures \citep{cen:99}. Low-frequency
   aperture arrays such as {\sc Lofar} are ideally suited to detect many
   diffuse steep-spectrum sources. In the next two years, the first {\sc
   Lofar} survey is expected to chart a million galaxies and may
   discover hundreds of cluster radio halos \citep{roettgering:06}. It
   is thus timely to study the distribution of diffuse radio sources in
   a cosmological context.
   \\

   There have been efforts to simulate the non-thermal emission from
   galaxy clusters by modelling discretised cosmic ray (CR) energy
   spectra on top of Eulerian grid-based cosmological simulations
   \citep{miniati:01b, miniati:04, miniati:07}. Recently, a series of
   papers explored the dynamical impact of CR protons on hydrodynamics
   in a cosmological SPH simulation \citep{jubelgas:08, pfrommer:06,
   ensslin:07}. \citet{skillman:08} presented a new method for
   identifying shock fronts in adaptive mesh refinement simulations and
   they computed the production of CRs in a cosmological volume adopting
   a nonlinear diffusive shock acceleration model. They found that CRs
   are dynamically important in galaxy clusters. \citet{pfrommer:08}
   used {\sc Gadget} simulations of a sample of galaxy clusters and
   implemented a formalism for CR physics on top of radiative
   hydrodynamics. They modelled relativistic electrons that are (i)
   accelerated at cosmological structure formation shocks and (ii)
   produced in hadronic interactions of CRs with protons of the ICM.
   \citet{pfrommer:08} approximated both the CR spectrum and that of
   relativistic electrons locally by single power-laws with free
   parameters for the slope, the normalisation, and the low energy
   cut-off. Energy and momentum conservation, including source and sink
   terms, result in evolution equations for the spectra. They found that
   the radio emission in galaxy clusters is dominated by secondary
   electrons. Only at the location of strong shocks the contribution of
   primary electrons may dominate. In the cluster centres they found a
   radio emission of about $10^2 \, h^3 \: {\rm mJy \, arcmin^{-2}}$,
   while in the periphery the azimuthal average the emission is $10^{-3}
   \, h^3 \: {\rm mJy \, arcmin^{-2}}$. 
   \\

   Little is still known about the structure of shock fronts in the ICM.
   With high resolution imaging one can constrain the width of the
   transition and a deconvolution of the images gives an estimate for
   the density and pressure jump. Since the mean free path of protons in
   the cluster environment is of the order of Mpc, shock fronts in the
   ICM are collisionless. As studies of the best investigated
   collisionless shock front, namely the Earth bow shock, have revealed,
   the dissipation of the upstream kinetic energy is a complex process
   and depends on several shock properties (see \eg\ \citet{burgess:07}
   for an introduction). For instance, the angle between the upstream
   magnetic field and the shock normal determines if an ion can gyrate
   between the upstream and downstream region, and the Mach number
   determines if instabilities in the shock region operate efficiently.
   However, shock fronts in the intra-cluster medium may differ
   significantly from the bow shock of the Earth as, for instance, the
   upstream plasma in the ICM is much less magnetised. Unfortunately, it
   is beyond the scope of current computer resources to study
   collisionless shock from the first principles, since scales from the
   electron gyro radius up to the large-scale structure of the shocks
   are involved. In hybrid simulations an effective small-scale response
   of electrons and protons is assumed. Using such a method,
   \citet{kang:07} found that upstream CRs excite Alfv{\'e}n waves and
   thereby amplify the magnetic field. 
   \\

   In this paper we combine a large cosmological simulation with a
   simple model for the radio emission of shock accelerated electrons.
   Our aim is to apply the emission model to the whole range of shock
   fronts generated during cosmic structure formation. A representative
   shock front sample is obtained from the \MNUniverse\ simulation which
   has been carried out with TreeSPH code {\sc Gadget}-2. We have
   developed a novel approach for locating the shock fronts and to
   estimate their Mach number. For computing the radio emission we
   follow Hoeft \& Br{\"uggen} (2007, HB07). They assumed that electrons
   are accelerated by diffusive shock acceleration and cool subsequently
   by synchrotron and inverse Compton losses. As a result the radio
   emission can be expressed as a function of downstream plasma
   properties, Mach number, and surface area of the shock front.
   Applying this radio emission model to the \MNUniverse\ simulation
   leads to radio-loud shock fronts with complex morphologies. We
   visualize these shock fronts to show where the radio emission is
   generated and compute artificial radio maps. As the \MNUniverse\
   simulation provides a cosmologically representative cluster sample,
   we also investigate the relation between radio luminosity and X-ray
   temperature. 
   \\

   This paper is organised as follows: In \Sec{sec-mn-universe} we
   briefly summarise the main characteristics of the \MNUniverse\
   simulation. In \Sec{sec-finding-shocks} we describe our approach for
   locating strong shock fronts in a SPH simulation and for estimating
   the Mach number. The radio emission model as worked out in HB07 is
   outlined and adopted for the SPH simulation in \Sec{sec-radio-model}.
   Finally, in \Sec{sec-results} we show our results for the shock
   fronts in the \MNUniverse\ simulation and for the radio properties of
   structure formation shocks.

\section{The Simulation}
\label{sec-mn-universe}

   To study the large-scale distribution of dark matter and gas in the
   universe, we have carried out a large cosmological gasdynamical
   simulations dubbed \MNUniverse\ \citep{gottloeber:07}. In this
   simulation, we assumed the spatially flat concordance cosmological
   model. In a series of lower resolution simulations we have studied
   the effect of different normalisation of the power spectrum
   \citep{yepes:07}.
   \\

   This paper is based on the original simulation with the cosmological
   parameters $\Omega_{\rm m} = 0.3$, $\Omega_{\rm bar} = 0.045$,
   $\Omega_{\Lambda} = 0.7$, the normalization $\sigma_8 = 0.9$ and the
   slope $n=1$ of the power spectrum. The simulation has been carried
   out with the {\sc Gadget}-2 code \citep{springel:05}. Within a box of
   $500\: h^{-1}\,{\rm Mpc}$ size the linear power spectrum at redshift
   $z=40$ has been represented by $1024^3$ dark matter particles of mass
   $m_{\rm DM} = 8.3 \times 10^{9} \; h^{-1} \, M_\odot $ and the same
   number of gas particles with mass $m_{\rm gas} = 1.5 \times 10^{9} \;
   h^{-1} \, M_\odot $. Within {\sc Gadget}-2 the equations of gas
   dynamics are solved by means of the smoothed particle hydrodynamics
   (SPH) method in its entropy conservation scheme. Radiative processes
   or star formation are not included in this simulation. The spatial
   force resolution was set to an equivalent Plummer gravitational
   softening of $15 \; h^{-1}\,{\rm kpc}$ (comoving) and the SPH
   smoothing length was set to the distance to the 40$^{th}$ nearest
   neighbor of each SPH particle.
   \\

   We have identified objects in this simulation using a parallel
   version of the hierarchical friends-of-friends (FOF) algorithm
   described in \citet{klypin:99}. The FOF algorithm is based on the
   minimal spanning tree (MST) of the particle distribution. The highest
   density peaks of the objects are calculated using a shorter linking
   length (1/8 of the virial linking length corresponding to 512 times
   the virial overdensity). The virial radius (overdensity 330) has been
   calculated around these centres. The most massive cluster contains
   about 250\,000 dark matter particles and 230\,000 gas particles
   within the virial radius. The mass of this cluster is $2.5 \times
   10^{15} \; h^{-1} \, M_\odot$. For the analysis presented in this
   paper, we selected the three hundred most massive clusters. The least
   massive cluster taken into account has a mass of $4 \times 10^{14} \;
   h^{-1} \, M_\odot$, corresponding to roughly 50\,000 dark matter and
   gas particles.

\section{Finding and characterising shock fronts in SPH simulations}
\label{sec-finding-shocks}

   The temperature of the intra-cluster medium (ICM), as derived from
   the X-ray emission, rises with the mass of a cluster. Roughly, the
   temperature is proportional to the velocity dispersion of the cluster
   galaxies, indicating that the gravitational energy gained during
   infall is mainly dissipated in the ICM. Since the latter is
   collisionless, meaning that Coulomb collisions are rare compared to
   the dynamical time scales, viscous dissipation is inefficient but
   shocks have to turn the energy gained by infall into heat. Therefore,
   shock heating has to be captured by any numerical scheme for
   simulating the formation of galaxy clusters.
   \\
   
   In smoothed-particle hydrodynamics (SPH) artificial viscosity is used
   to dissipate energy in shocks. Shock fronts are detected by
   evaluating the velocity field within the SPH kernel. Negative
   velocity divergence indicates a region of a shock. The form of the
   artificial viscosity is tuned to obey the jump conditions of
   discontinuities and to avoid spurious dissipation at shear flows.
   Simulations of the Sod shock tube problem \citep{landau:59} and
   comparison with the analytic solution show that artificial viscosity
   reliably determines the dissipation at a shock front, see \eg\ 
   \citet{springel:05}. However, this formalism does not rely on
   macroscopic properties of shock fronts, such as the density or
   temperature jump. Since we will need the Mach number of the shock to
   compute the radio emission, we have to locate the shock fronts in the
   simulation and to determine their Mach number.

\subsection{Hydrodynamical shocks}
\label{sec-non-rad-shock}

   In cosmic gas flows the bulk velocity often exceeds the local sound
   speed. As a result shock fronts develop and dissipate the kinetic
   energy. The front separates the pre-shock ({\it upstream}) and the
   post-shock ({\it downstream}) regime. At the shock front the coherent
   motion of the upstream gas is partly randomised, meaning that part of
   the kinetic energy is converted into heat. Dissipation may procced
   via an intricate sequence of processes, however, mass, momentum, and
   total energy fluxes are conserved, except for radiative losses.  For
   non-radiative, unmagnetised shocks, upstream and downstream density,
   $\rho$, velocity, $v$, pressure, $P$, and specific internal energy,
   $u$, are related by
   \begin{eqnarray}
	 \rho\up v\up
	 &=&
	 \rho\down v\down
	 \nonumber
	 \\
	 P\up +  \rho\up v\up^2
	 &=&
	 P\down +  \rho\down v\down^2
	 \label{eq-conservation}
	 \\
	 \frac{1}{2}v\up^2 + u\up + \frac{P\up}{\rho\up}
	 &=&
	 \frac{1}{2}v\down^2 + u\down + \frac{P\down}{\rho\down}
	 \nonumber
	 ,
   \end{eqnarray}
   where the velocities have to be measured in the rest-frame of the
   shock surface. For a more detailed description of jump conditions,
   see, \eg\ \citet{landau:59}. The entropy, in contrast, increases at
   the shock discontinuity due to the dissipative processes. As a
   measure for entropy we will use for simplicity the entropic index
   \begin{equation}
	 S 
	 =
	 u \,
	 \rho^{1-\gamma} 
	 ,
	 \label{eq-def-entr-index}
   \end{equation}
   where $\gamma$ denotes the adiabatic exponent. For given upstream
   plasma conditions, a shock front is entirely characterised by the
   upstream Mach number
   \begin{equation}
	 {\cal M}
	 =
	 \frac{v\up}{c\up}
	 ,
	 \label{eq-def-mach}
   \end{equation}
   where $c\up$ denotes the upstream sound speed, $c\up^2 = \gamma
   (\gamma-1) u\up$. Alternatively, the shock front can also be
   characterised by the compression ratio,
   \begin{equation}
	 r 
	 =
	 \frac{\rho\down}{\rho\up}
	 ,
	 \label{eq-def-compress-ratio}
   \end{equation}
   or the entropy ratio,
   \begin{equation}
	 q 
	 =
	 \frac{S\down}{S\up}
	 .
	 \label{eq-def-entr-ratio}
   \end{equation}
   In order to identify shock fronts in the \MNUniverse\ simulation and
   to determine the related radio emission, it will be useful to express
   the Mach number and the compression ratio as a function of the
   entropy ratio. Therefore, we relate these three properties by the
   help of the definition of the Mach number, \Eq{eq-def-mach}, and the
   conservation equations, \Eqs{eq-conservation}. For the Mach number we
   find
   \begin{equation}
	 {\cal M}^2
	 =
	 \frac{1}{c\up^2} \:
	 \frac{\rho\down}{\rho\up} \:
	 \frac{P\down - P\up}{\rho\down - \rho\up}
	 .
	 \nonumber
   \end{equation}
   Furthermore, assuming that the plasma obeys the polytropic relation,
   \ie\ $P=(\gamma-1)\rho u$, we obtain
   \begin{equation}
	 {\cal M}^2
	 =
	 \frac{r}{\gamma} \: 
	 \frac{ q r^\gamma - 1 }{ r - 1 }
	 .
	 \label{eq-mach-from-q-r}  
   \end{equation}
   For a polytropic plasma the compression ratio and entropy ratio
   are related by the implicit equation 
   \begin{equation}
	 r
	 =
	 \frac{ (\gamma+1) \, q \, r^\gamma + (\gamma-1)}
		  { (\gamma-1) \, q \, r^\gamma + (\gamma+1)}
	 .
	 \label{eq-r-from-q}
   \end{equation}
   \Eqs{eq-mach-from-q-r} and (\ref{eq-r-from-q}) provide a one-to-one
   mapping between the Mach number, compression ratio, and entropy ratio.
   Finally, the difference between downstream and upstream velocity is
   related to the Mach number via
   \begin{equation}
	 v\down - v\up
	 =
	 {\cal M} \,
	 \frac{r-1}{r} \,
	 u\up
	 .
   \end{equation}
   Our Mach number estimator computes the entropy ratio, $q$, and
   $(v\down - v\up)/u\up$ for each SPH particle in the simulation.
   Beforehand, we tabulate the relations between $q$, $r$, ${\cal M}$
   and ${\cal M}(r-1)/r$. This allows us to simply read the Mach number
   from the table. The implementation of the Mach number estimator is
   explained in \Sec{sec-estimate-mach}.
   \\
    
   Most likely, the structure of cosmological shock fronts is more
   complex than that of simple hydrodynamical shocks. It is beyond the
   scope of this paper to include a more detailed structure of
   collisionless shocks. Instead, we combine the shock fronts detected
   in the simulation with a parametric model for the radio emission. Any
   aspect of the shock structure has to be captured by the parameters
   used for the radio emission since cosmological simulations are not
   able to resolve any relevant scale for plasma processes.

\subsection{A novel approach for estimating the Mach number}
\label{sec-estimate-mach}

    \begin{figure}
	 \begin{center}
	 
	 \includegraphics[width=0.45\textwidth,angle=0]{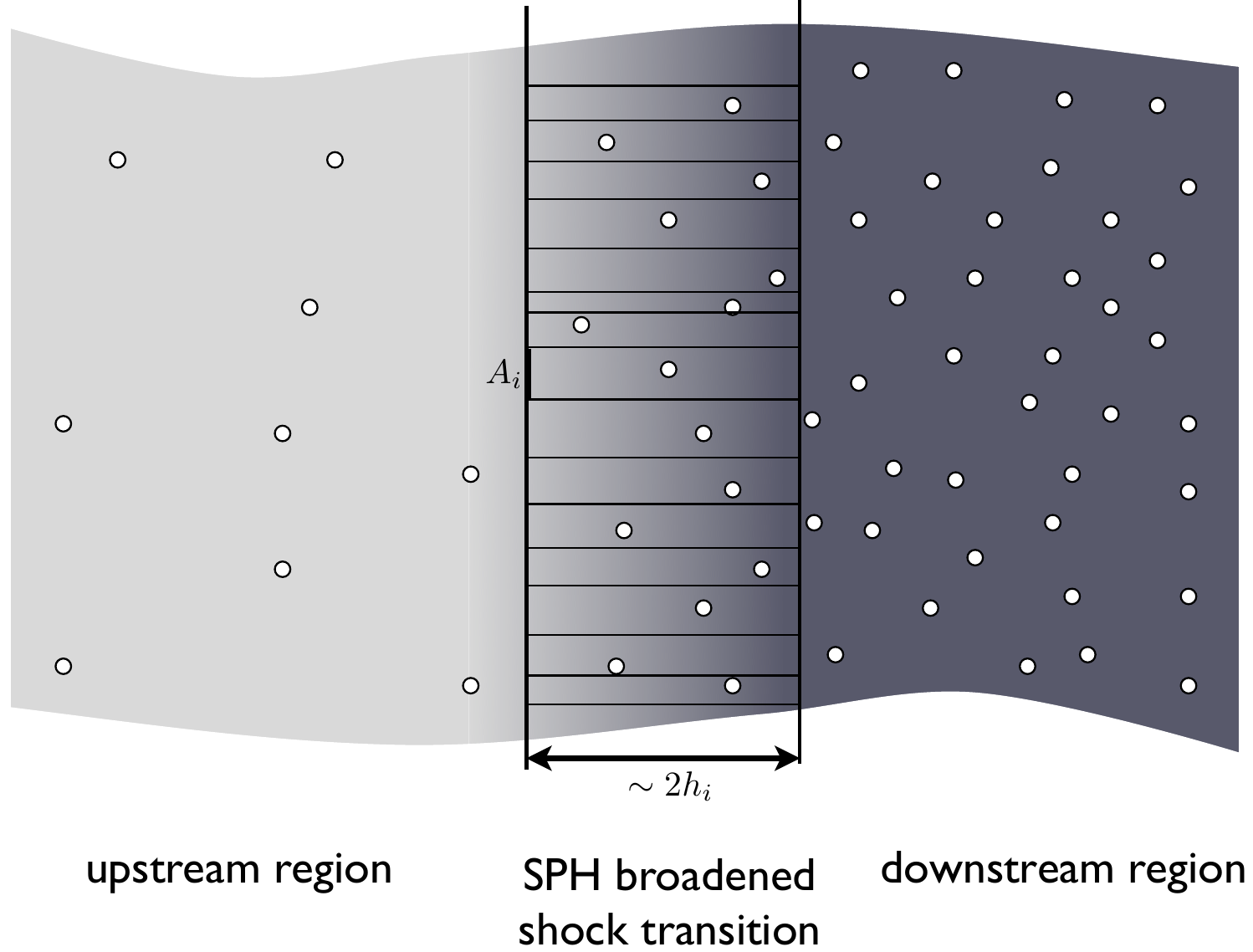}

	 \end{center}
	 \caption
	 { 
     The shock structure in SPH simulations.
	 }
	 \label{fig-shock-surface-element}
   \end{figure}

   Our aim is to identify shock fronts in the \MNUniverse\ simulation
   and to derive the Mach number of the shocks. In order to estimate the
   radio emission produced by the shock, we will need the Mach number,
   downstream temperature and electron density. In SPH, energy is
   dissipated at shocks using an artificial viscosity. Most
   implementations of artificial viscosity follow the suggestions by
   \citet{monaghan:92} and \citet{balsara:95}. In addition,
   viscosity-limiters are used to reduce spurious angular momentum
   transport in shear flows \citep{balsara:95,steinmetz:96}.  In
   contrast to actually solving a Riemann problem, the SPH formalism
   does not identify any discontinuity. For our purpose it is sufficient
   to identify the shock fronts by post-processing snapshots of the
   simulation.
   \\
   
   One possible approach is to extract the ratio of the upstream and
   downstream entropy. In SPH the dissipation rate is computed but the
   upstream and downstream entropy are never computed. The ratio of
   these two quantities would immediately give the strength of the
   shock. \citet{pfrommer:06} determine the entropy gain in the shock by
   computing the product of the dissipation rate and the estimated
   crossing time of particles through the smoothed shock front. However,
   for particles in the SPH broadened shock front the analysis of local
   properties underestimates the strength of a shock.
   \citet{pfrommer:06} cure this difficulty by using the maximum Mach
   number from within a certain time interval. As we wish to determine
   the Mach number from a single snapshot, we have developed the
   following scheme: 
   \\
   
   In the first step, for each SPH particle, we compute the entropy
   gradient, $\nabla S$. In a shock front this gradient gives the direction
   of the shock normal pointing into the downstream direction. We define
   an associated upstream position,
   \begin{equation}
	 \vec{x}_{{\rm u}i}
	 =
	 \vec{x}_i 
	 + 
	 f_h \, h_i \,\vec{n}_i^1
	 ,
	 \label{eq-upstream-position}
   \end{equation}
   where $\vec{x}_i$ is the position of the SPH particle $i$, $h_i$ the
   smoothing length of the particle $i$, and $\vec{n}_i^1$ denotes the
   shock normal, $\vec{n}_i^1 = - \nabla S / |\nabla S|$. Moreover, we
   have introduced a factor $f_h$, chosen to be 1.3, which ensures
   that $\vec{x}_{{\rm u}i}$ is indeed in the upstream region and not in the
   transition between upstream and downstream properties. In a similar
   manner, we introduce an associated downstream position, $\vec{x}_{{\rm
   d}i}$, in the opposite direction.
   \\

   Using the usual SPH scheme, we compute the density and internal
   energy at the upstream and downstream position. Moreover, we compute
   the velocity at these two positions and project it onto the shock
   normal. This results in the upstream and downstream velocity,
   $v_{{\rm u}i} + v_{\rm sh} = \vec{v}(\vec{x}_{{\rm u}i}) \cdot
   \vec{n}_i^1$ and $v_{{\rm d}i} + v_{\rm sh} = \vec{v}(\vec{x}_{{\rm
   d}i}) \cdot \vec{n}_i^1$, where $v_{\rm sh}$ is the velocity of the
   shock front in the rest-frame of the simulation. Since $v_{\rm sh}$
   is not known, it is advantageous to assess only the differences $(v_{{\rm
   d}i} - v_{{\rm u}i})$. Beside a component parallel to the shock
   normal, the velocity field may also show perpendicular components. We
   compute these components in a similar manner to the upstream and
   downstream velocities, $v_{i}^{k\pm} = \vec{v}( \vec{x}_i \pm f_h h_i
   \vec{n}_{i}^k ) \cdot \vec{n}_{i}^{k} $, where the three vectors
   $\vec{n}^1$, $\vec{n}^2$, and $\vec{n}^3$ form an orthonormal base.
   \\

 For those particles, that belong clearly to a
   shock front, we demand that the velocity has to
   be divergent in the direction of the shock normal, \ie\ $(v\down -
   v\up)>0$. Moreover, the velocity differences in the directions
   perpendicular to the shock normal have to be smaller than that
   parallel to the shock, $|v_k^+ - v_k^- | < ( v\down - v\up ) / 2$.
   Therefore, the difference $( v\down - v\up )$ gives roughly the
   velocity divergence.
   \\

   There are two possibilities to compute the Mach number: First, we
   can evaluate the entropy ratio $q=S(\vec{x}\down)/S(\vec{x}\up)$.
   Secondly, we can use the ratio $(v\down - v\up)/c\up$. As discussed in
   \Sec{sec-non-rad-shock}, for both properties there exists a one-to-one
   mapping with the Mach number of the shock. We use a look-up table to
   obtain for both the entropy ratio and the velocity difference the
   corresponding Mach number. In order to have a conservative estimate
   for the Mach number we use the smaller of the two.

\section{The radio emission model in a nutshell}
\label{sec-radio-model}

   At collisionless shock fronts a small fraction of particles may be
   accelerated to energies far beyond the thermal energy distribution.
   This has been directly observed in particle spectra of the Earth's bow
   shock where the resulting momentum distribution depends strongly on
   the Mach number and the angle between shock normal and the direction
   of the upstream magnetic field. Also, the radio emission of solar
   bursts is generally attributed to shock-drift acceleration and
   subsequent excitation of plasma wave. Larger electron energies may be
   achieved by diffusive shock acceleration (DSA). The details of
   particle acceleration at collisionless shocks are complex and not
   fully understood yet. It is expected that DSA can operate in the ICM
   only with a starting energy above a few MeV, since at lower energies Coulomb
   losses are too efficient. Some electrons may attain this injection
   energy by shock-drift acceleration. It is beyond the scope of this
   paper to include a very detailed model for the processes at
   collisionless shock fronts. For simplicity, we assume that DSA
   operates already at thermal energies where Coulomb losses have to be
   compensated by acceleration mechanisms different from DSA. Following
   our model \citet[HB07]{hoeft:07} we assume that supra-thermal
   electrons show a power-law spectrum, with a slope as predicted by
   DSA. We also suppose that a fixed fraction of the energy dissipated
   at the shock front is used to accelerate electrons. Finally, we
   assume that the supra-thermal electrons advect with the downstream
   plasma. In this section, we briefly introduce diffusive shock
   acceleration and review the model discussed in HB07. Moreover, we
   discuss our assumptions about the magnetic field. Finally, we
   present shock tube simulations that test our shock finder and allows
   us to calibrate the emission model.

\subsection{Diffusive shock acceleration}
\label{sec-dsa}

   In DSA, particles are accelerated by multiple shock crossings, in a
   first-order Fermi process. If the shock thickness is much smaller
   than the diffusion scale which, in turn, has to be much smaller than
   the curvature of the shock front, a one-dimensional
   diffusion-convection equation can be solved
   \citep{axford:78,bell:78a,bell:78b,blandford:78}. The result is that
   the energy spectrum of suprathermal electrons is a power-law
   distribution, $n_E \propto E^{-s}$. The spectral index, $s$, of the
   accelerated particles is only related to the compression ratio at the
   shock front
   \begin{equation}
	 s
	 =
	 \frac{r+2}{r-1}
	 .
	\label{eq-s-from-r}  
   \end{equation}
   For strong, non-radiative shocks with Mach number $\gtrsim$10, the
   compression ratio is always close to 4, hence the slope $s$ is always
   close to 2. This can explain the spectrum of cosmic rays over a
   huge range of energies and may be considered a piece of evidence
   for DSA. For a review of diffuse shock acceleration see
   \citet{drury:83,blandford:87,jones:91,malkov:01}. 
   \\

   In supernova remnants, there is evidence that electrons and protons
   are accelerated by diffuse shock acceleration to energies of
   $\rm\sim10^{15}\,eV$ \citep{berezhko:03,vink:03}.

\subsection{Total emission induced by a shock front}
\label{sec-total-radio}

   In HB07 we have developed a model for the synchrotron
   emission from electrons that have been accelerated by DSA. Here,
   we summarise briefly the model and adapt it for the use in the
   framework of a SPH simulation. Our basic assumption is that a small
   fraction, $\xi_{\rm e}$, of the energy dissipated at the shock front
   is used to accelerate electrons to relativistic energies.
   The theory of diffusive shock acceleration predicts that the
   supra-thermal electron distribution obeys a power-law,
   \begin{equation}  
	 \frac{{\rm d}n_{\rm e}}
		  {{\rm d}E}
	 =
	 C_{\rm spec} \,
	 f_{\rm spec} ( e, s )
	 =
	 C_{\rm spec} \,
	 n_{\rm e}
	 \frac{ 1 }{ m_{\rm e} c^2} \,
	 e^{-s}
	 \qquad
	 {\rm for }
	 \qquad
	 e > e_{\rm min} 
	 ,
	 \label{eq-e-supr-spec} 
   \end{equation}
   where $n_{\rm e}$ is the electron density and $C_{\rm spec}$ is a
   normalisation factor. The kinetic energy is measured in units of the
   electron rest mass, $e = E / m_{\rm e} c^2 $. The energy $e_{\rm
   min}$ denotes the transition energy from the thermal
   Maxwell-Boltzmann distribution to the supra-thermal power-law
   distribution. In HB07 we have determined the normalisation factor,
   \begin{equation} 
	 C_{\rm spec}
	 =
	 \xi_e
	 \frac{u\down}{c^2}
	 \frac{m_{\rm p}}{m_{\rm e}} 
	 \frac{(q-1)}{q}
	 \frac{1}{F_{\rm spec}( e_{\rm min}, s )}
	 ,
	 \label{eq-C1}
   \end{equation}
   where $F_{\rm spec}( e_{\rm min}, s )$ is the integral of the
   power-law distribution without normalisation,
    \begin{equation} 
	 F_{\rm spec}( e_{\rm min}, s )
	 =
	 \int\nolimits_{e_{\rm min}}^\infty \, 
	 {\rm d}e'
	 f(e',s)
	 .
   \end{equation}
   The transition energy, $e_{\rm min}$, has to be derived iteratively
   from the condition that supra-thermal electrons carry the fraction
   $\xi_{\rm e}$ of the kinetic energy.
   \\

   For computing the radio emission we assume furthermore that the
   electrons are advected with the downstream plasma. This is justified
   since the electron gyro radii are small, orders of magnitude below
   one pc for any reasonable magnetic field strength and electron
   energy. Downstream electrons with $e\gtrsim 100$ cool mainly by
   synchrotron and inverse Compton losses \citep{sarazin:99}. These
   electrons also contribute most to the detectable radio emission. For
   these two cool mechanisms the evolution of electron distribution
   function can be described analytically \citep{kardashev:62},
   \begin{equation}
     \frac{{\rm d}n_{\rm e}}
		  {{\rm d}E}
	 (t)
	 =
	 C_{\rm spec} \,
	 n_{\rm e} \,
	 \frac{1}{m_{\rm e}c^2} \:
	 e^{-s}
	 \left\{ 
	   1 -  C_{\rm cool} \, e  \, t 
	 \right\}^{s-2} 
	  ,
	 \label{eq-e-spectrum-solution} 
   \end{equation}
   with 
   \begin{equation}
	 C_{\rm cool}
	 =
	 \frac{4 \, \sigma_{\rm T}}{3 \, m_{\rm e} \, c} \, \{ u_{\rm CMB} + u_B \} 
	 .
	 \label{eq-cooling-constant} 
   \end{equation} 
   Here  $u_{\rm CMB}$ denotes the energy density of the cosmic
   microwave background, $u_B$ the energy density of the downstream
   magnetic field, and $\sigma_{\rm T}$ the Thomson cross section. We
   assume that in the zone relevant for radio emission the electron
   density and the magnetic field strength are constant. Evidently,
   the ICM varies in the region of the radio relics. We still ignore
   this effect since much larger uncertainties derive from our poor
   knowledge about the microphysics of collisionless shocks in the ICM.
   \\

   The radio emissivity, $P(e,\nu_{\rm obs})$, depends on the energy of
   the electron energy, $e$, the observing frequency, $\nu_{\rm obs}$,
   and the magnetic field strength, $B$, \citep{rybicki:86}. Convolving
   the emissivity with the electron distribution,
   \Eq{eq-e-spectrum-solution}, and integrating over the downstream
   region results in the total emission behind the shock front,
   \begin{equation}  
	 \frac{{\rm d} P ( \nu_{\rm obs} )}{{\rm d} \nu}
	 =
	 m_{\rm e} c^2 \,
	 A \,	 
	 \int\nolimits_0^\infty {\rm d } y \;
	 \int_{E_{\rm min}}^\infty  {\rm d }E \;
	 \frac{{\rm d}n_{\rm e}}
		{{\rm d}E}
	 (t(y)) \:
	 P(e,\nu_{\rm obs})
	 ,
	 \label{eq-total-emission}
   \end{equation}
   where $y$ denotes the normal distance of a downstream location to the
   shock front, and $A$ is the surface area of the shock front. In HB07
   we have evaluated the integral, \Eq{eq-total-emission}. We found that
   the emission  can be given by the shock surface area, the Mach
   number, the downstream temperature, the electron density, and the
   magnetic field strength,
   \begin{eqnarray}  
	 \frac{{\rm d} P ( \nu_{\rm obs} )}{{\rm d} \nu}
	 & = &
	 6.4 \times 10^{34} \, {\rm erg \, s^{-1} \, Hz^{-1} } \;\;
	 \frac{A}{{\rm Mpc^2}} \,
	 \frac{ n_{\rm e}}{\rm 10^{-4} cm^{-3}} \,
         \nonumber
	 \\
	  && \quad
     	 \times
	 \frac{ \xi_{\rm e} }{0.05 } \:
	 \left(\frac{\nu_{\rm obs}}{\rm 1.4 \, GHz} \right)^{-\frac{s}{2}}
	 \left( \frac{T\down}{\rm 7\, keV} \right)^{\frac{3}{2}} \:
	 \label{eq-total}
	 \\
	 && \quad
	 \times
	 \frac{ ( B / {\rm  \mu G})^{1+\frac{s}{2}} }
		   {     (B_{\rm CMB} / {\rm \mu G} )^2
			   + ( B / {\rm \mu G} )^2 }
	 \;
	 \Psi({\cal M}, T\down )
	 .
	 \nonumber
   \end{eqnarray}
   The function $\Psi({\cal M}, T\down )$ comprises all dependencies on
   the Mach number. Basically, $\Psi({\cal M}, T\down )$ rises steeply
   for Mach numbers in the range 2 - 4 and is constant for large Mach
   numbers.
   \\

   In HB07 we integrated \Eq{eq-total-emission} for a steady-state
   situation, where all time stages of the electron distribution function
   are present, from the newly accelerated distribution to a long time
   cooled spectrum. In a steady-state situation the overall synchrotron
   emission spectrum obeys a power-law. The slope of the spectrum is
   $s/2$, even though, locally, the slope may vary from the slope of newly
   accelerated electrons, $(s-1)/2$, to a very steep spectrum. This
   distribution of the spectral index agrees well with the
   observation for several radio relics (\eg\ \citet{orru:07,pizzo:08}).
   \\

   As discussed in HB07 the integral, \Eq{eq-total-emission}, can also
   be used to estimate the extent in $y$-direction, normal to the shock.
   We found that the extent depends mainly on the magnetic field. In the
   cluster centre the strength may be of the order of $1\,\mu {\rm G}$.
   In this case, 50\,\% of the radio emission comes from a region with a
   size of a few hundred kpc in $y$-direction, \ie\ the size of the
   radio object is of the order of the size of the cluster centre. In
   contrast, in the cluster  periphery the magnetic field strength may
   be of the order of $0.1\,\mu {\rm G}$. In this case, the extent of
   the emission region is about 100\,kpc, which is significantly smaller
   than the scale of spatial variations in the cluster periphery. In the
   following, we assign the radio emission solely to the shock front and
   ignore the extent of the radio emission region in $y$-direction. This
   approach is perfectly suited for studying the radio emission in the
   cluster periphery, since there the downstream extend of the emission
   region is smaller than the numerical resolution. In the cluster
   centre our approach may lead to too compact radio objects.
   \\

   The shock finder described in \Sec{sec-estimate-mach} allows us to
   locate shock discontinuities in a SPH simulation snapshot. The shock
   finder also provides estimates for the Mach number and for the shock
   normal. Using the normal vector, we can find a position which is
   sufficiently downstream to determine the downstream temperature,
   $T\down$, and the electron density, $n_{\rm e}$, which are needed to
   compute the radio emission via \Eq{eq-total}. To compute the radio
   power, ${\rm d} P / {\rm d} \nu$, we also need the area, $A$, of the
   shock front. Since in SPH a shock front is comprised of particles,
   each particle represents an area of the shock. More precisely, the
   shock discontinuity is smoothed by the size of the SPH kernel,
   $h_{\rm SPH}$, see \Fig{fig-shock-surface-element}. The kernel may
   contain $N_{\rm SPH}$ particles, the corresponding volume is of the
   order of $h_{\rm SPH}^3$. Hence, one particle belonging to the shock
   front represents a shock area of
   \begin{equation}
	 A
	 =
	 f_A
	 \frac{  h_{{\rm SPH}}^2 } 
		  { N_{\rm SPH } }
	 ,
   \end{equation}
   where $f_A$ is a normalisation constant, which will be determined by
   the help of Sod shock tube simulations, see \Sec{sec-shock-tube}.
   Finally, we have to determine the downstream magnetic field strength.
   In the next paragraph, \Sec{sec-magn-field} we detail how we estimate
   $B\down$.

\subsection{Magnetic field model}
\label{sec-magn-field}

   The downstream magnetic field is crucial for the synchrotron
   emission. For a power-law electron energy distribution the emission
   is given by 
   \begin{equation}
     P(\nu)
     \propto
     n_{\rm e} \,
     ( B \sin\theta )^{1+\alpha} \,
     \nu^{-\alpha}
     ,
     \label{eq-powerlaw-emission}
   \end{equation}
   where $\theta$ is the pitch angle between the electron velocity and
   the magnetic field. The exponent $\alpha$ is related to the slope of
   the electron spectrum by $\alpha = (s-1)/2$. Unfortunately, in
   \Eq{eq-powerlaw-emission} the electron density and the field strength
   are degenerated. Therefore, evaluating only the synchrotron emission
   of diffuse radio objects in the ICM is not sufficient to derive the
   magnetic field strength. For some clusters it has been possible to
   measure the electron density by the hard X-ray excess in the cluster
   spectrum (see \citet{rephaeli:08} for a review). As a result, the
   average cluster magnetic field strength has been estimated to be of
   the order of $0.1$ to $1\,{\rm \mu G}$ \citep{fusco-femiano:01,
   rephaeli:03, rephaeli:06}. Alternatively, one assumes that the energy
   density of the ICM is minimal \citep{govoni:04}. This method also
   results in magnetic field strengths of the order of $0.1$ to $1\,{\rm
   \mu G}$. Contrarily, Faraday rotation measurement studies (see \eg\
   \citet{govoni:06}) typically yield field strengths that are larger by
   about an order of magnitude.  
   \\

   The strength, structure and origin of magnetic fields in the ICM is
   still debated. Intergalactic magnetic fields may be a remainder of
   primordial processes, may be generated by a dynamo in first stars and
   subsequently expelled, may be generated by the Biermann battery
   effect during cosmic reionisation, may be caused by
   microinstabilities in the turbulent ICM, or may be generated at
   collisionless shock fronts. It is beyond the scope of this paper to
   include a detailed model for the generation and evolution of magnetic
   fields. Even the passive evolution of magnetic fields during cosmic
   structure formation is complex \citep{dolag:02,brueggen:05}. However,
   we assume here that the magnetic strength in the downstream region of
   the shock front, $B\down$, on average simply obeys flux conservation
   to follow
   \begin{equation}
      \frac{ B\down }
           { B_{\rm ref } }
      =
      \left(
        \frac{ n\down } 
             { 10^{-4} \, {\rm cm^{-3} } }
      \right)^{2/3}
      .
      \label{eq-B-dens-rel}
   \end{equation}

\subsection{Normalisation of the radio emission}
\label{sec-shock-tube}

   \begin{figure*}
	 \begin{center}
	 \includegraphics[width=0.9\textwidth]{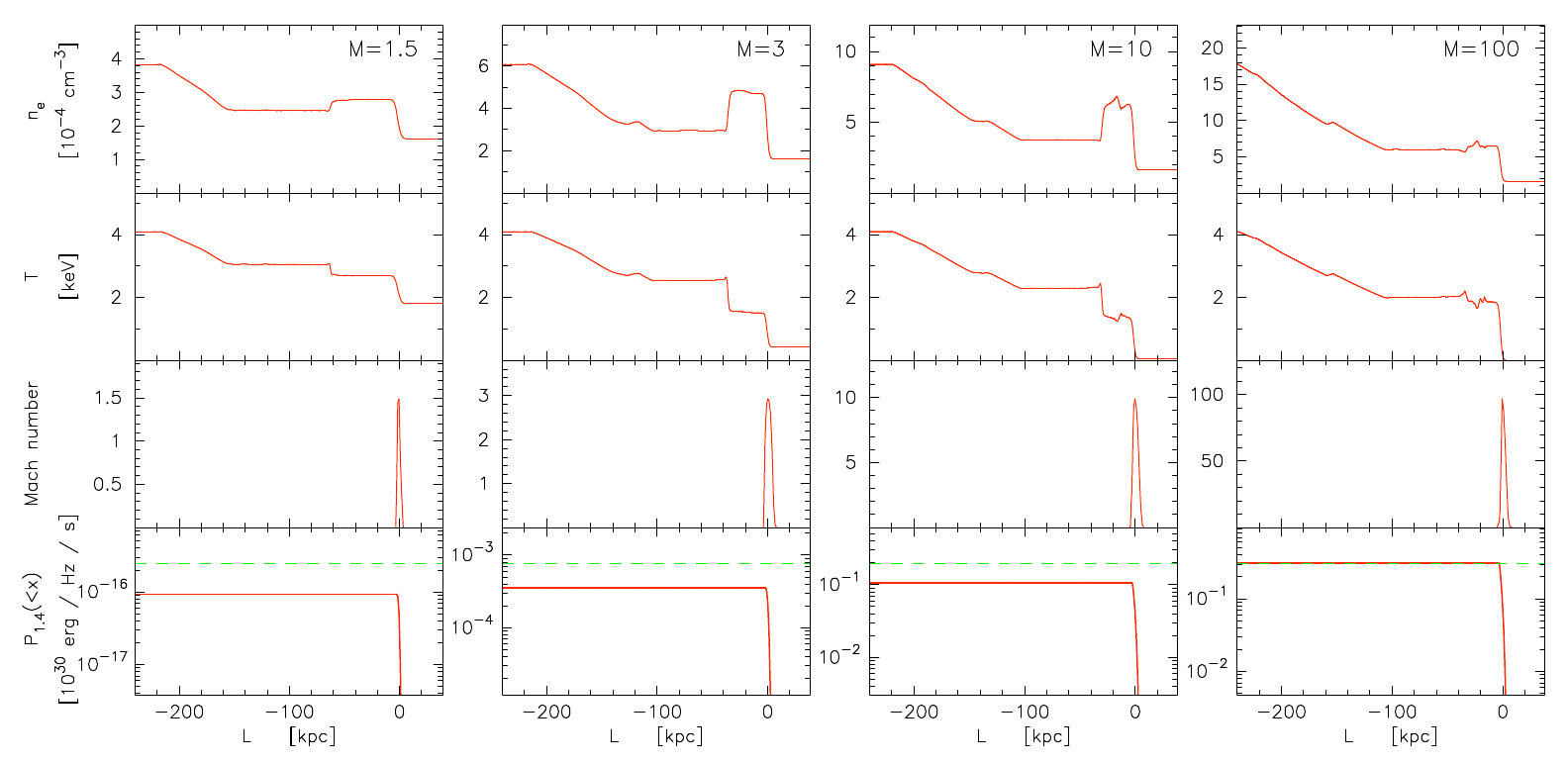}
	 \end{center}
	 \caption
	 { 
	 Simulations of the Sod shock tube problem with Mach numbers 1.5, 3,
	 10, 100. The upper panel shows the electron density assuming a
	 fully ionised ICM, the second row the temperature, the third row
	 the Mach number as determined by the Mach number estimator
	 described in \Sec{sec-estimate-mach} and the lower panel the
	 cumulative radio emission as described in \Sec{sec-shock-tube}
	 assuming an observation frequency of 1.4\,GHz. Note that the
	 irregularites in $n_{\rm e}$ and $T$ are caused by the small number
	 of particles in the SPH kernel, namely $N_{\rm SPH} = 40$. We
	 choose this number in order to conform with the parameters used in
	 the \MNUniverse\ simulation. For the shock finding we smoothed the
	 fields to some extend by using $N_{\rm kernel} = 80$ in the finder,
	 see \Tab{tab-para}.  
	 }
	 \label{fig-shock-tube}
   \end{figure*}

   In our model it is necessary to determine the effective area of the
   shock front represented by one SPH particle, $A_i$, which may also
   depend on the actual Mach number. Therefore, we simulated the Sod
   shock tube problem and determine the Mach number by our method
   described in \Sec{sec-estimate-mach}. As shown in
   \Fig{fig-shock-tube}, the original Mach number used to set up the
   initial conditions is well reproduced by the Mach number estimator.
   We have simulated shock tubes with Mach numbers from 1.5 to 100. This
   covers well the range of relevant Mach numbers for the radio emission
   of merger-induced and accretion shocks. In all cases, the
   theoretically expected Mach number is found in the centre of the
   transition region between the upstream and the downstream region. In
   the periphery the Mach number is underestimated, \ie\ the particles
   in the shock region show a Mach number distribution, as shown in the
   third row of \Fig{fig-shock-tube}.
   \\

   The theoretical value of the radio emission is easily computed via
   \Eq{eq-total}. The Mach number is known from setting up the initial
   conditions of a shock tube, and the properties of the downstream
   plasma are known from the analytic solution. In the simulation a
   large number of SPH particles contributes to the emission. To show
   the total emission we plot the cumulative radio emission, \ie\ the
   emission, $P(x_i)$, of all particles on the left-hand side of a given
   $x$-position is summed,
   \begin{equation} 
     P(<x) 
     = 
     \sum\nolimits_i \,
     P(x_i) \qquad
     {\rm with } \qquad
     x_i < x
     .
     \label{eq-radio-cum}
   \end{equation}
   As noted above the Mach number estimator results in a distribution of
   Mach numbers for the particles in the SPH-broadened shock
   discontinuity. For small Mach numbers, ${\cal M}\lesssim 5$, the
   radio emission depends strongly on ${\cal M}$, while for large Mach
   numbers it does not. Hence, a constant area normalisation factor,
   $f_A$ (introduced in \Sec{sec-total-radio}), would systematically
   underestimate the radio emission of shocks with ${\cal M}\lesssim 5$.
   One could correct for this effect by introducing a function
   $f_A({\cal M})$. However, we choose a somewhat simpler solution and
   increase the Mach number slightly when computing the radio emission
   for a particle,
   \begin{equation}
     \Psi({\cal M})
     \to
     \Psi({\cal M}\times 1.045)
     .
   \end{equation}
   As a result we are able to reproduce the expected radio emission of
   the different shock tubes with deviations smaller than a factor of
   three. Regarding the simplicity of our model (power-law electron
   spectrum, uniform energy fraction in supra-thermal electrons, flux
   conservation of magnetic field) this accuracy is sufficient.

   \begin{table*}
     \centering
     \begin{tabular}{ l l c c c }
     Module  & Parameter & Symbol & Value  & See Sec. for details  \\
     \hline
     \hline
        \MNUniverse  
           & mass dark matter part. 
           & $m_{\rm DM}$ 
           & $8.3 \times 10^{9} \: h^{-1} \, \Msun $ 
           & \ref{sec-mn-universe} 
        \\
           & mass gas particles 
           & $m_{\rm gas}$
           & $1.5 \times 10^{9} \: h^{-1} \, \Msun $
           & 
        \\
           & force resolution 
           & $r_{\rm soft}$
           & $15 \: h^{-1} \, {\rm kpc }$ (comoving)
           & 
        \\
           & SPH kernel 
           & $N_{\rm SPH}$
           & 40
           & 
        \\
     \hline
        Shock finder  
           & shock extent factor 
           & $f_h$ 
           &  1.3  
           & \ref{sec-estimate-mach} 
        \\
           & part. number in kernel 
           & $N_{\rm kernel}$ 
           & 80 
           &  \ref{sec-estimate-mach} 
        \\
     \hline
        Mach number estimator  
           & shock area fraction 
           & $f_A$ 
           & 6.5 
           & \ref{sec-total-radio}, \ref{sec-shock-tube} 
        \\
           & Mach number correction 
           & ${\cal M }$
           & ${\cal M} \times 1.045 $ 
           & \ref{sec-shock-tube} 
        \\
     \hline
        Radio emission 
           & energy fract. in supra-thermal e$^-$
           & $\xi_{\rm e}$ 
           & $5\times10^{-3}$ 
           & \ref{sec-total-radio} 
        \\
           & reference magn. field strength  
           & $B_{\rm ref}$ 
           & $0.1\muGauss$ 
           & \ref{sec-magn-field} 
        \\
     \hline
     \end{tabular}
     \caption{
     Parameters used in the \MNUniverse\ simulation, for finding shock
     fronts, for estimating the Mach number, and for assigning the radio
     emission.
     }
     \label{tab-para}
   \end{table*}

\section{The radio emission in the simulation}
\label{sec-results}

   We now wish to apply the tools prepared in the previous sections to
   estimate the radio emission of strong shocks which occur during
   cosmic structure formation. The most luminous diffuse radio emission
   is observed in massive clusters or their periphery. We therefore
   analyse massive clusters from a cosmological simulation. Our aim is
   to apply the radio emission model to the whole range of structure
   formation shocks, caused by mergers and gas accretion. In this paper
   we estimate only the radio emission which is originated by primary
   electrons. 
   \\

   We apply the Mach number estimator described in
   \Sec{sec-estimate-mach} and the radio emission model described in
   \Sec{sec-total-radio} to the \MNUniverse\ simulation described in
   \Sec{sec-mn-universe}. More precisely, we analyse the 300 most
   massive clusters with masses in the range 0.4 to $2.5 \times 10^{15}
   h^{-1} \Msun $. We do not include less massive clusters since the
   resolution gets too poor. First we show the results of our Mach
   number estimator for a selection of clusters. Then, we present
   artificial observations and compare them to observed radio relics.
   Finally, we discuss how the radio emission correlates with X-ray
   temperature of the clusters.

\subsection{The ICM in slices}
\label{sec-slices}

   \begin{figure*}
	 \begin{center}

	 \includegraphics[width=0.32\textwidth,angle=0]{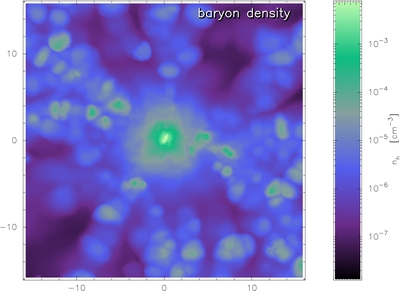}
	 \includegraphics[width=0.32\textwidth,angle=0]{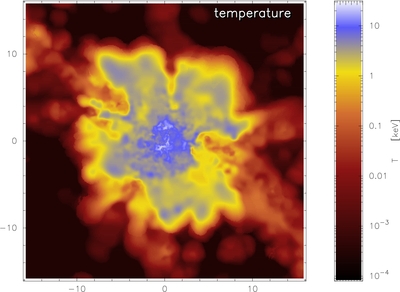}
	 \includegraphics[width=0.32\textwidth,angle=0]{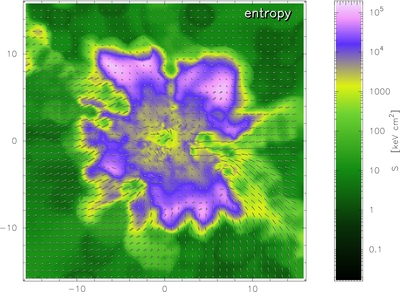}
	 
	 \includegraphics[width=0.32\textwidth,angle=0]{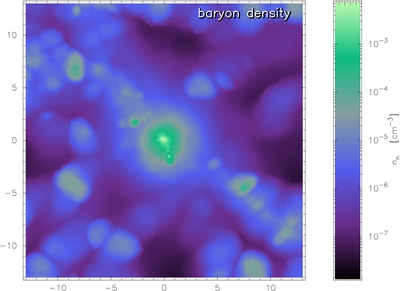}
	 \includegraphics[width=0.32\textwidth,angle=0]{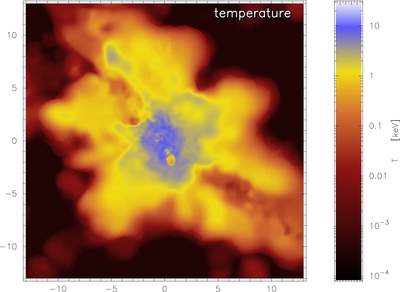}
	 \includegraphics[width=0.32\textwidth,angle=0]{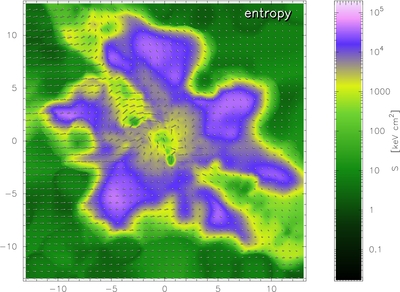}
	 
	 \end{center}
	 \caption
	 { 
	 Density, temperature and entropy in a slice for two clusters. Each
	 row represents one cluster. Density, temperature and entropy are
	 computed for a slice through the cluster centre, by computing for
	 each pixel the SPH average. Spatial scales are given in $h^{-1}
	 \:{\rm Mpc}$. The virial radius on the two clusters is about $2 \:
	 h^{-1} \:{\rm Mpc}$. One can nicely see that gas gets heated at a
	 few times the virial radius.
	 }
	 \label{fig-slices}
   \end{figure*}

   \begin{figure*}
	 \begin{center}
	 
	 \includegraphics[width=0.95\textwidth,angle=0]{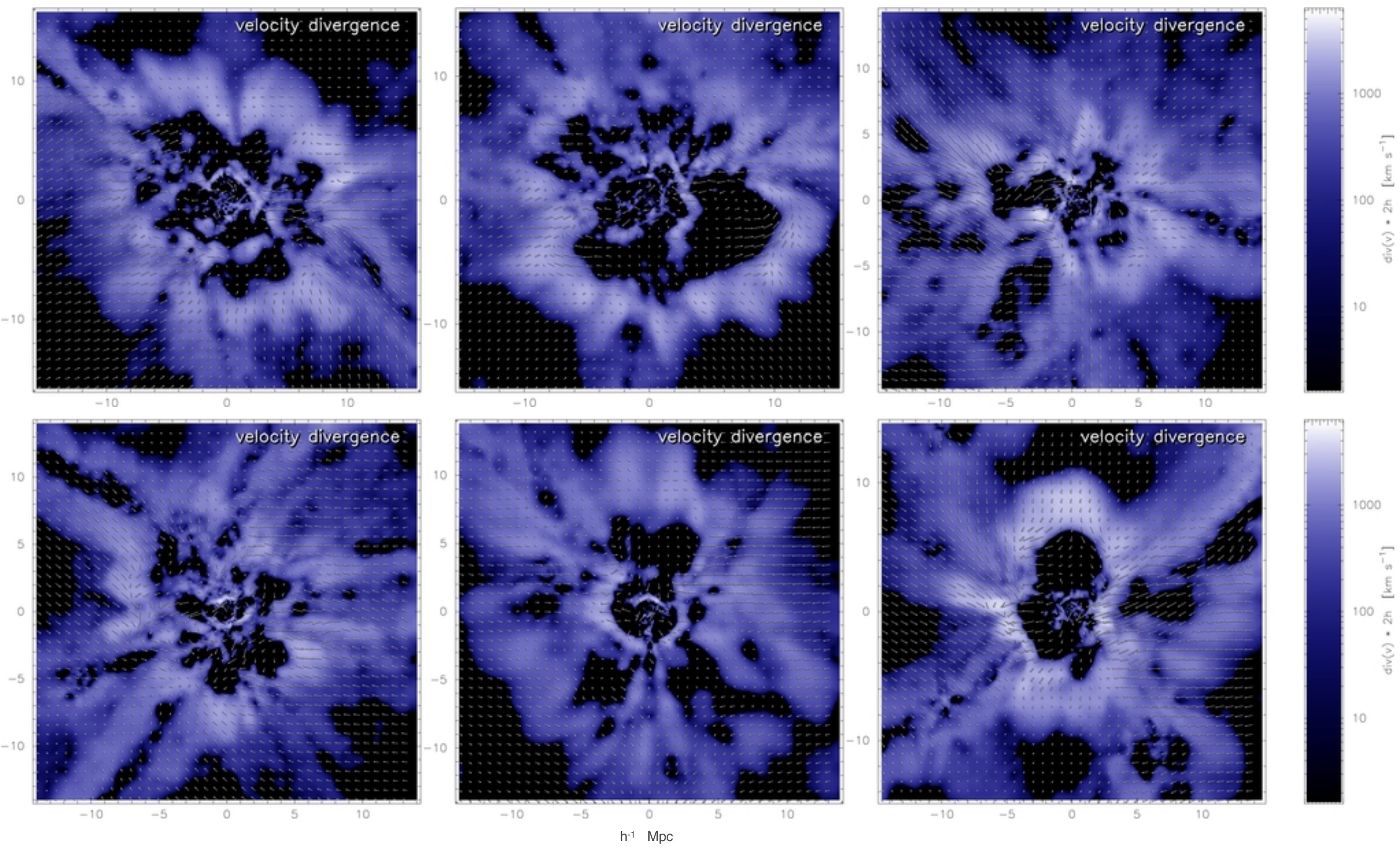}

	 \end{center}
	 \caption
	 { 
	 Velocity divergence in a slice through the cluster center for six
	 different clusters of our sample. The velocity divergence refers
	 here to the difference $(v\down -v\up)$, see
	 \Sec{sec-finding-shocks}. One can clearly distinguish the outer
	 `accretion' shocks and the inner `merger' shocks. Note that the
	 width of the shocks is caused by the the numerical resolution,
	 which is much poorer in the region of the accretion shocks than in
	 inner region of the clusters.
	 }
	 \label{fig-vel-divergence}
   \end{figure*}

   It is instructive to consider the physical properties of the ICM in
   slices through the cluster, see \Fig{fig-slices}. One can clearly see
   how the temperature rises sharply in the periphery of a cluster, far
   beyond the virial radius. At this location, about four times the
   viral radius, the ICM gets heated to cluster tempertaures, \ie\ about
   $1\,{\rm keV}$, by accretion shocks. Within clusters, the temperature
   and entropy maps are complex, since rather cold intergalactic medium
   flows in filaments to the cluster centre and ongoing or past mergers
   lead to intricate gas flows. However, the velocity divergence reveals
   also the location of the inner merger shocks, see
   \Fig{fig-vel-divergence}. Note that we compute the divergence
   according to the procedure described in \Sec{sec-finding-shocks},
   \ie\ we determine first the shock normal and subsequently compute the
   difference $(v\down-v\up)$. As also described in
   \Sec{sec-finding-shocks} the ratio of the velocity divergence and the
   upstream sound speed allows us to infer the Mach number of the shock.
   Additionally, the ratio of downstream and upstream entropy allows to
   determine the Mach number. We use the minimum of both for further
   analysis, \ie\ for computing radio emission.

\subsection{Visualising the shock fronts}
\label{sec-viz-fronts}

   \begin{figure*}
	 \begin{center}
	 
	 \includegraphics[width=0.8\textwidth,angle=0]{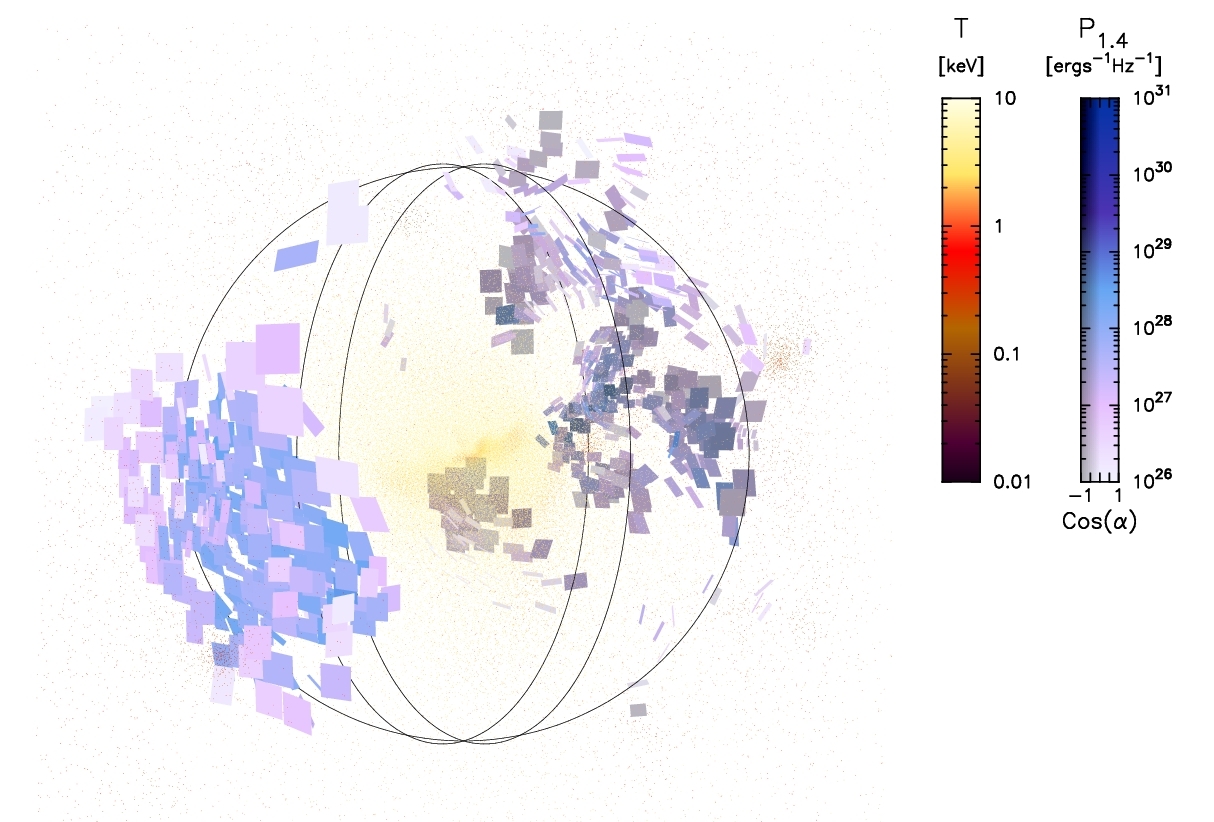}

	 \end{center}
	 \caption
	 { 
     A textbook example of a double relic found in the \MNUniverse\ simulation.
	 }
	 \label{fig-textbook-relic}
   \end{figure*}

   \begin{figure*}
	 \begin{center}
	 
	 \includegraphics[width=0.85\textwidth,angle=0]{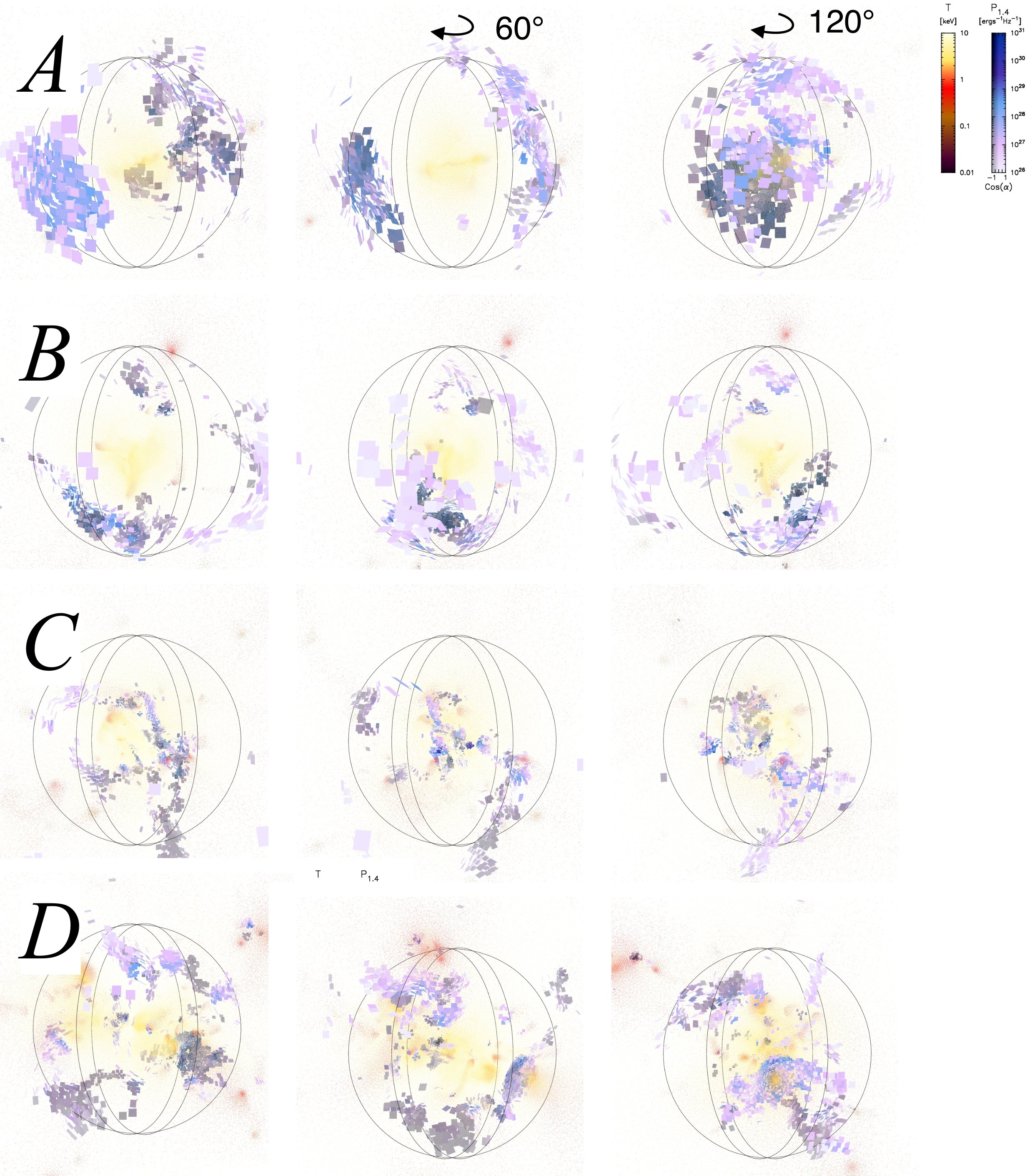}
	
	 \end{center}
	 \caption
	 { 
	 Radio emission and particle distribution for four typical galaxy
	 clusters. The particles of the simulation are represented by dots
	 with temperature colour-coded. Each SPH particles which belongs to
	 a shock front a small square is drawn. The size of the square is
	 proportional to the smoothing length of the particle. The colour
	 gives the radio emission associated with the particle. Moreover we
	 encode if a square is seen from upstream or the downstream
	 direction. The latter are shaded grey. The circles indicate the
	 viral radius of the galaxy cluster. See \Tab{tab-clusters} for the
	 properties of these exemplaric clusters. Each row represents one
	 cluster, rotated by azimuthal angles of $60^\circ$ and $120^\circ$.
	 See \Fig{fig-radio-projection} for the projected radio emission.
	 }
	 \label{fig-radio}
   \end{figure*}

   \begin{table*}
     \centering
     \begin{tabular}{ c c c c c c c l }
       Label  
         & $L_{\rm bol}$ 
         & $T_X$ 
         & $M_{\rm vir}$  
         & $t_{\rm merger}$
         & $P_{1.4}(<2R_{\rm vir})$  
         & $\langle B\down \rangle $
         & morphology has
       \\
         & $[ 10^{44} \, \rm erg \, s^{-1} ]$ 
         & $[\rm keV]$  
         & $[10^{15} \, h^{-1} \, M_\odot ]$  
         & $[{\rm Gyr}] $
         & $[ 10^{31} \rm erg \, s^{-1} \, Hz^{-1} ]$  
         & $[{\rm \mu  G}]$
         & similarities to
       \\
     \hline
      A 
        & \,\, 2.8
        & \,\, 3.7
        &      0.7  
        &      2.6
        & \,\, 0.35
        &      0.07	
        & A3667, A3376
      \\
      B 
        &  \,\, 6.4
        &  \,\, 4.8
        &       0.9
        &       3.0
        &  \,\, 0.81
        &       0.12
        & A115, A1664
      \\
      C 
        &       50.4
        &       12.2
        &        1.5
        &        3.1
        &  \,\,  8.56
        &        0.79
        & A520
      \\
      D 
        &     11.5
        & \,\, 4.1
        &      1.9
        &      2.0
        &     11.52
        &      0.49
        &
      \\[.2ex]
      \hline
      E 
        &      19.0
        &  \,\, 7.8
        &       1.4
        &       5.3
        &  \,\, 0.04
        &       0.25
        & 
      \\
      F 
        &  \,\,  7.6
        &  \,\,  5.3
        &        1.2
        &        4.1
        &  \,\, 0.12
        &       0.07
        &
      \\
      G 
        &       23.4
        &   \,\, 8.1
        &        1.1
        &        1.9
        &   \,\, 0.10
        &        0.12
        & 
      \\
      H 
        &  \,\, 5.0
        &  \,\, 3.7
        &       0.8
        &       1.2
        &  \,\, 0.03
        &       0.16
        &
      \\
     \hline
     \hline
     \end{tabular}
     \caption{
     Properteis of the selected clusters. 
     }
     \label{tab-clusters}
   \end{table*}

   As seen in the slices through the clusters, the geometry of the
   strong shocks may be very complex. To visualise the shocks, we
   represent each of these particles by a small square to which the
   shock normal is perpendicular. Each of these square may be considered
   as a small part of the shock surface. The edge length of a square is
   proportional to the smoothing length of the corresponding SPH
   particle.
   \\

   We color-code the squares by the radio emission computed for the
   corresponding SPH particle. \Fig{fig-textbook-relic} shows a
   text-book example of a double relic found in the \MNUniverse\
   simulation. This cluster has developed two large shock fronts in the
   cluster periphery, similar to the radio objects found in A3667
   \citep{roettgering:97} and A3376 \citep{bagchi:06}. In addition to the
   emission of each SPH particle in the shock front, we color-code the
   orientation of the shock normal. We shade squares grey 
   for which the normal points away from the observer, where the
   level of shading depends on the angle, $\alpha$, between
   shock normal and the direction to the observer. As a result, shock
   fronts which propagate towards the observer appear blueish while
   fronts which move away appear greyish.
   \\

   To show the location of the cluster, we represent the SPH particles
   by dots with the temperature colour-coded.  Infalling substructures
   can often be identified in temperature maps since they are usually
   much colder than the surrounding ICM. The clusters in the
   \MNUniverse\ simulation show radio objects with very different
   morphologies. In \Fig{fig-radio} four examples are displayed, each of
   which viewed from three different angles. As discussed above, cluster
   {\em A} contains a textbook example for a double relic. Cluster {\em
   B} shows only one relic which is composed of several smaller
   fragments. In cluster {\em C}, several strongly emitting sources in
   the centre of the cluster are found. Finally, cluster {\em D} shows
   the emission of an early state merger system. The temperature of this
   cluster is only about 4 keV (see \Tab{tab-clusters}), but still a
   strong radio relic has been developed. 
   \\

   The structure of the shock fronts can be seen best in an animation.
   We therefore encourage the reader to view the movies of the rotations
   of the four exemplary clusters available from our web site
   \footnote{\tt www.faculty.iu-bremen.de/mhoeft/RadioRelicAnimation/}.

\subsection{Radio maps}
\label{sec-maps}

    \begin{figure*}
	 \begin{center}
	 
	 \includegraphics[width=0.9\textwidth,angle=0]{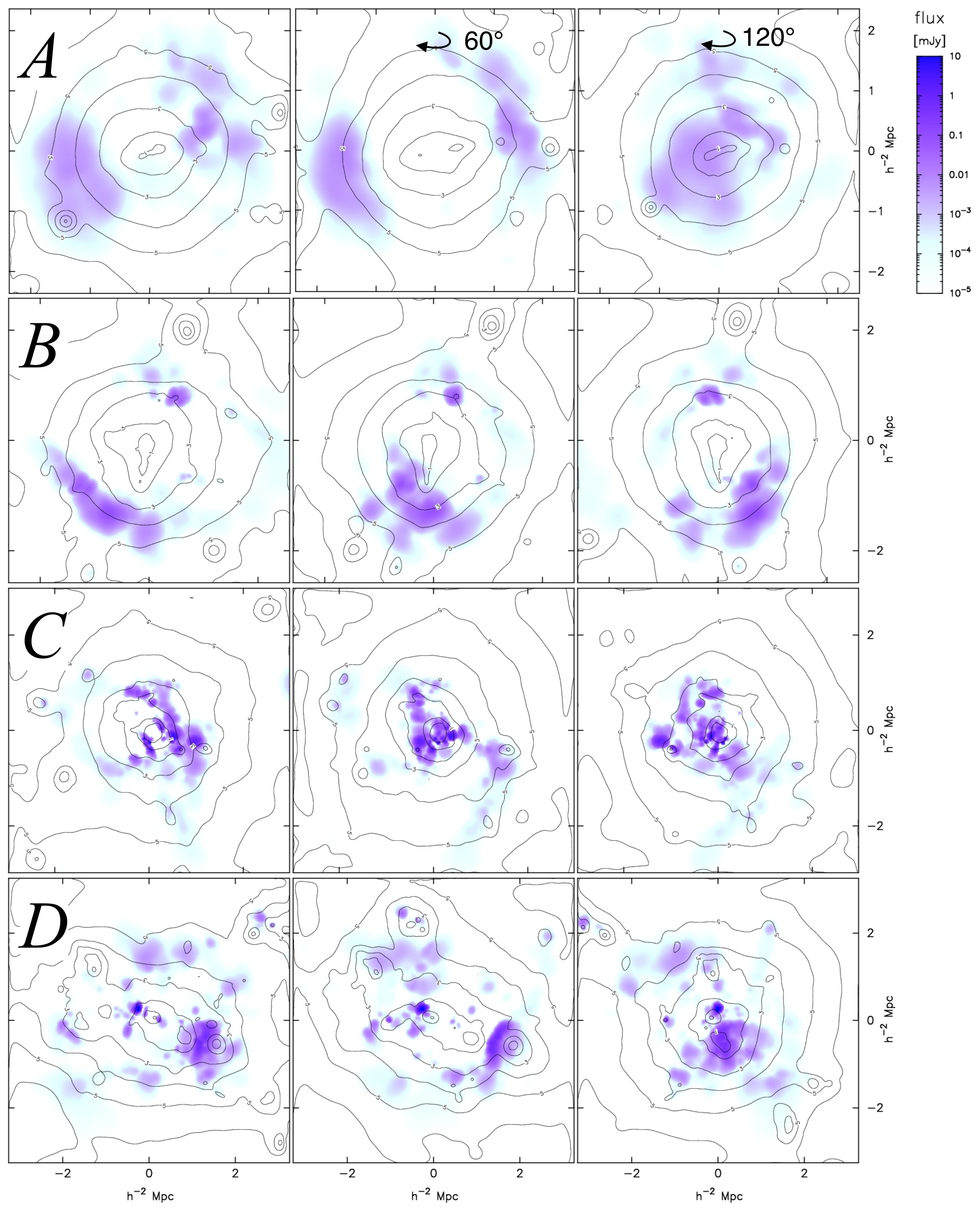}
	 
	 \end{center}
	 \caption
	 { 
	 Mimic observations of the X-ray and the radio emission. We have
	 projected the radio emission and bolometric X-ray emission of the
	 clusters shown in \Fig{fig-radio}. The radio flux at $1.4\,{\rm
	 GHz}$ is computed for a hypothetical beam size of $10'' \times
	 10''$. No redshift effects are taken into account. 
	 }
	 \label{fig-radio-projection}
   \end{figure*}

   Having assigned a radio luminosity to each SPH particle, we are able
   to compute artificial radio maps. To this end we project the emission
   of each particle along the line-of-sight and smooth it according to
   the SPH kernel size. We compute the radio emission for an observing
   frequency of 1.4\,GHz and a hypothetical beam size of $10'' \times
   10''$. For comparison, we also compute contours of the bolometric
   X-ray flux. We follow \citet{navarro:95} and assign each particle
   the luminosity
   \begin{equation}
     L_X
     =
     1.2 \times 10^{24} \, {\rm erg \, s^{-1} } \:
     \frac{m_{\rm gas} }{ \mu \, m_{\rm p } } \:
     \frac{ n_{\rm e} } { \rm cm^{-3} }
     \left( \frac{T}{\rm keV } \right)^{1/2}
     .
   \end{equation}  
   \Fig{fig-radio-projection} shows the artificial radio maps and the
   X-ray contours for the same clusters already presented in
   \Fig{fig-radio}. The double relic structure of cluster {\em A}
   becomes even more prominent. The X-ray contours reveal for some
   viewing angles clearly the merger state of the clusters.
   \\

   The examples in \Fig{fig-radio-projection} and \Fig{fig-radio} are
   selected to illustrate the variety of diffuse radio objects obtained
   by our simple emission model. Example {\em A} shows the large-scale
   merger shocks that are also found in idealised merger simulations
   \citep[e.g.][]{roettiger:99}. In contrast, example {\em C} shows a
   large number of smaller shock fronts. Some of them are related to
   substructures  moving through the ICM, while others are obviously
   unrelated to any substructure (this can be best seen in the
   animations mentioned above). The morphology of the radio emission
   resembles that found in A2255, where several rather small pieces of
   diffuse radio emission have been detected \citep{pizzo:08}.
   \\

\begin{figure*}
	 \begin{center}
	 
	 \includegraphics[width=0.95\textwidth,angle=0]{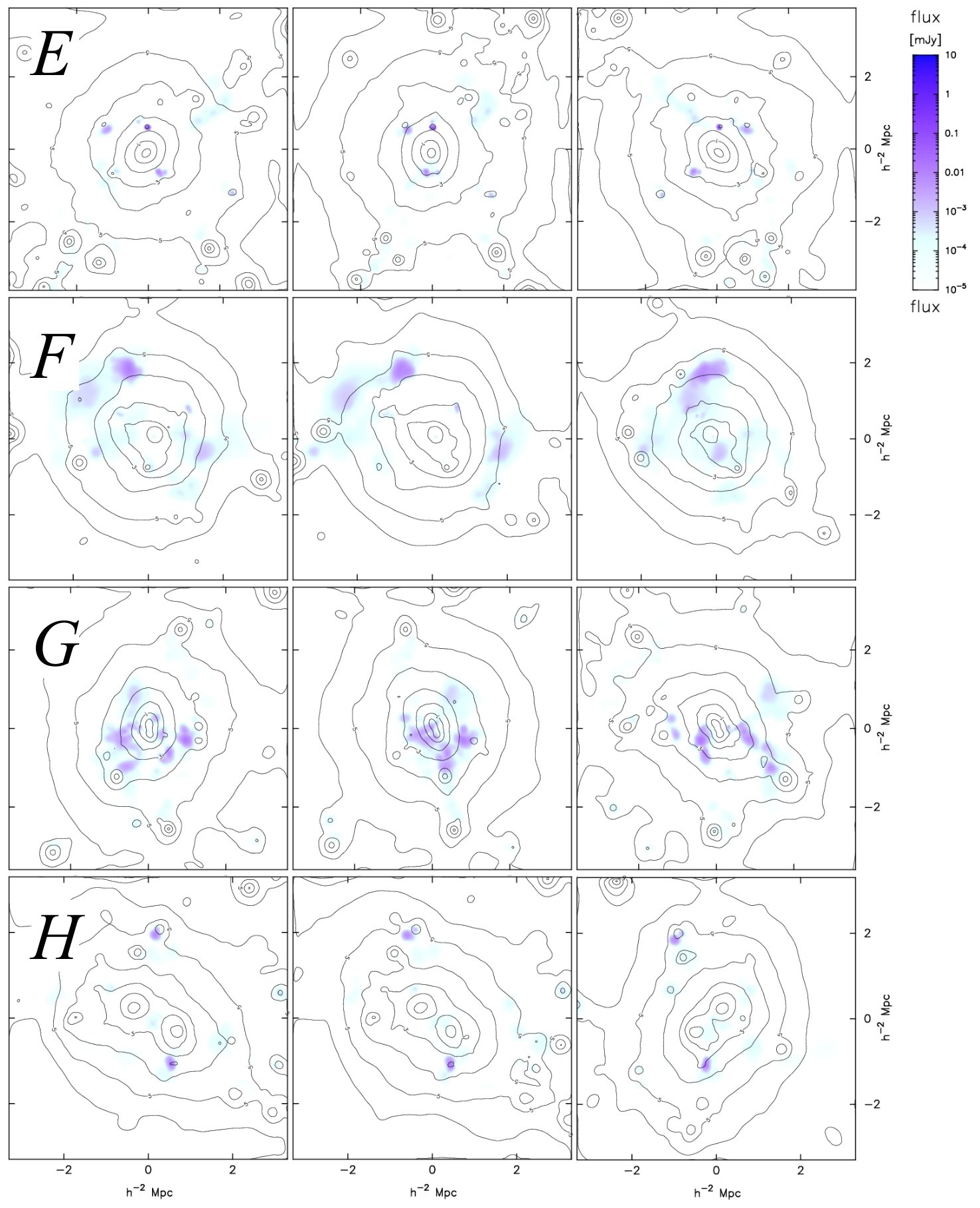}
	
	 \end{center}
	 \caption
	 { 
	 Mimic observations of the X-ray and the radio emission for four
	 galaxy clusters with a low radio emission. 
	 }
	 \label{fig-radio-little}
   \end{figure*}

   Another property of our model for diffuse radio emission is the fact
   that many of the massive clusters show no or only little radio
   emission. Therefore, we show four examples with almost radio-quiet
   clusters, see \Fig{fig-radio-little}. Example {\em E} is the
   stereotype of a relaxed massive clusters. There is no large-scale
   shock present which can cause significant radio emission, only a very
   few small regions with a radio-loud shock are found. This is
   consistent with the time of the last major merger (the mass of the
   accreted substructure has to be at least 25\,\% of the main cluster),
   which has take place more than $5\,Gyr$ in the past. Examples {\em D}
   and {\em F} show a distorted X-ray morphology indicating recent
   merger activity. However, the resulting merger shocks are very faint
   in the radio. Finally, example {\em G} shows a merger in an early
   state with no radio emission. Hence, even with our simple radio
   emission model merger activity does not lead in all cases to strong
   radio emission. This is expected from modelling electron acceleration
   at the shock front, since a Mach number above 2.5 is needed for a
   significant amount of radio emission. Many of the merger shocks have
   lower Mach numbers.

\subsection{The luminosity function of radio objects}
\label{sec-radio-LF}

   \begin{figure}
	 \begin{center}
	 \includegraphics[width=0.45\textwidth,angle=0]{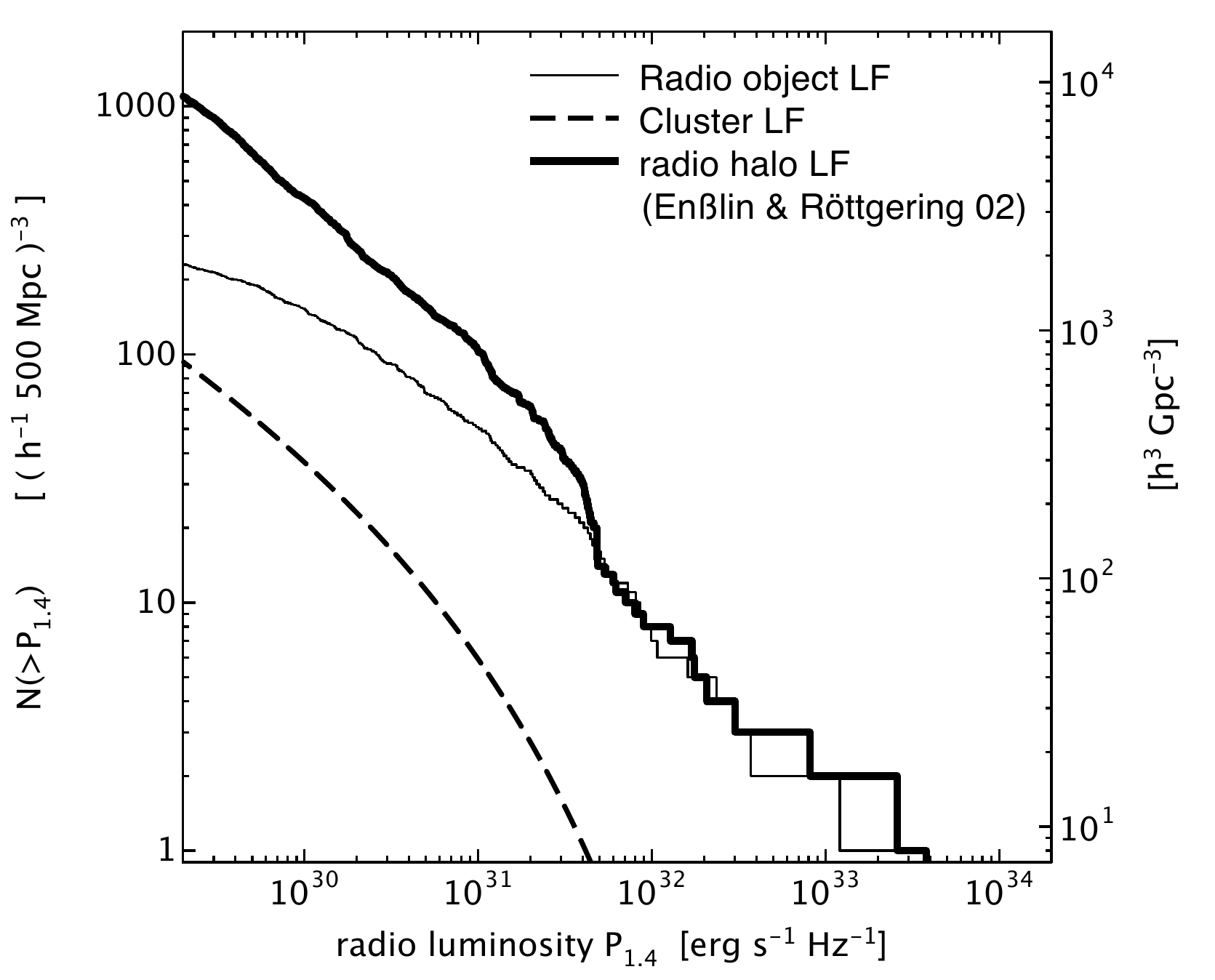}
	 \end{center}
	 \caption
	 { 
	 The cumulative number density of radio objects. The left axis give
	 the number of objects found in the simulation box. The solid line
	 indicates the number of radio objects as identified by splitting
	 the objects in well separated groups (see \Sec{sec-radio-LF}). The
	 long-dashed line gives the radio luminosity function for the
	 clusters in \MNUniverse\ simulation. The radio power of a cluster
	 includes all emission within the virial radius. For comparison we
	 depict the radio halo luminosity function as computed by
	 \citet{ensslin:02b}. We adopt their best fitting of the radio
	 halo--X-ray correlation with $a_\nu = 3.37$ and $b_\nu = 1.69$.
	 }
	 \label{fig-cum-synchro}
   \end{figure}

   As can be seen by the examples {\em A} to {\em D}, one can often
   divide the emitting structures into individual objects, such as the
   emission in front of a supersonically moving galaxy and that of a
   merger shock front travelling outwards in the ICM. In order to
   identify individual radio objects in our simulation, we select all
   particles whose radio emission lies above a very low threshold. For
   these particles we determine their nearest neighbours, and obtain
   as a result a spanning tree for radio-loud particles. We now cut
   all links that are larger than the minimum of the smoothing lengths
   of the two connected particles. As a result, the spanning tree
   separates into sub-trees, each of which can be considered an
   individual radio object. By this method we can for instance
   disentangle the radio emission of an infalling substructure from
   that of large merger shock. Finally, we compute the cumulative
   number of radio objects above a given luminosity, see
   \Fig{fig-cum-synchro}, \ie\ the luminosity function of diffuse
   radio objects. Most of the very luminous objects ($P_{1.4} \gtrsim
   10^{32} \, {\rm erg \, s^{-1} \, Hz^{-1} }$) are rather small
   structures, similar to that seen in example {\em C}. The textbook
   merger shocks, as in example {\em A}, have only luminosities of the
   order of $10^{31} \, {\rm erg \, s^{-1} \, Hz^{-1} }$ and below.
   Therefore, the very luminous relics in A3667 seem to be very
   exceptional. Our textbook merger shocks are similar to the less
   luminous relics found in A3376 \citep{bagchi:06}. 
   \\

   We choose the parameters of our model in a way that the resulting
   radio objects agree with observations in the respect that the most
   luminous ones have a radio power, $P_{1.4}$, of $\sim 10^{33} \, {\rm
   erg \, s^{-1} \, Hz^{-1} }$. Towards small luminosities the
   cumulative number of radio objects increases with a slope close to 
   $-2/3$. For comparison we compute the radio luminosity within the
   virial radius of the clusters in the \MNUniverse\ simulation and find
   an even shallower curve. The rather shallow slope of the faint end of
   the radio luminosity function reflects that the radio luminosity
   depends strongly on the cluster mass or temperature. We
   will discuss this further in the next section.
   \\

   For comparison we have shown in \Fig{fig-cum-synchro} the radio halo
   luminosity derived by \citet{ensslin:02b}. They concluded that the
   upcoming radio telescope {\sc Lofar} could discover about 1000 radio
   halos by a survey in a years timescale. With the parameters used in
   our model we find a much larger abundance of radio objects. However,
   we may overestimate the efficiency of diffusive shock acceleration or
   the strength of magnetic fields. It is worth noticing that the slope
   of the radio luminosity function at the low-luminosity end is
   similar to that of \citet{ensslin:02b}, hence, one may expect the
   discovery of a large number of radio objects with {\sc Lofar}. The
   radio luminosity function of clusters is significantly flatter,
   indicating that one cluster may contain several radio objects with
   moderate luminosity. As a result we expect that the number of cluster
   for which {\sc Lofar} will find new radio features is significantly
   lower than 1000. However, for a more quantitative analysis model
   parameters have to be normalised and the redshift evolution has to be
   included properly. This will be discussed in a forthcoming paper.

\subsection{Radio versus X-ray} 
\label{sec-radio-xray}

   \begin{figure}
	 \begin{center}
	 \includegraphics[width=0.45\textwidth,angle=0]{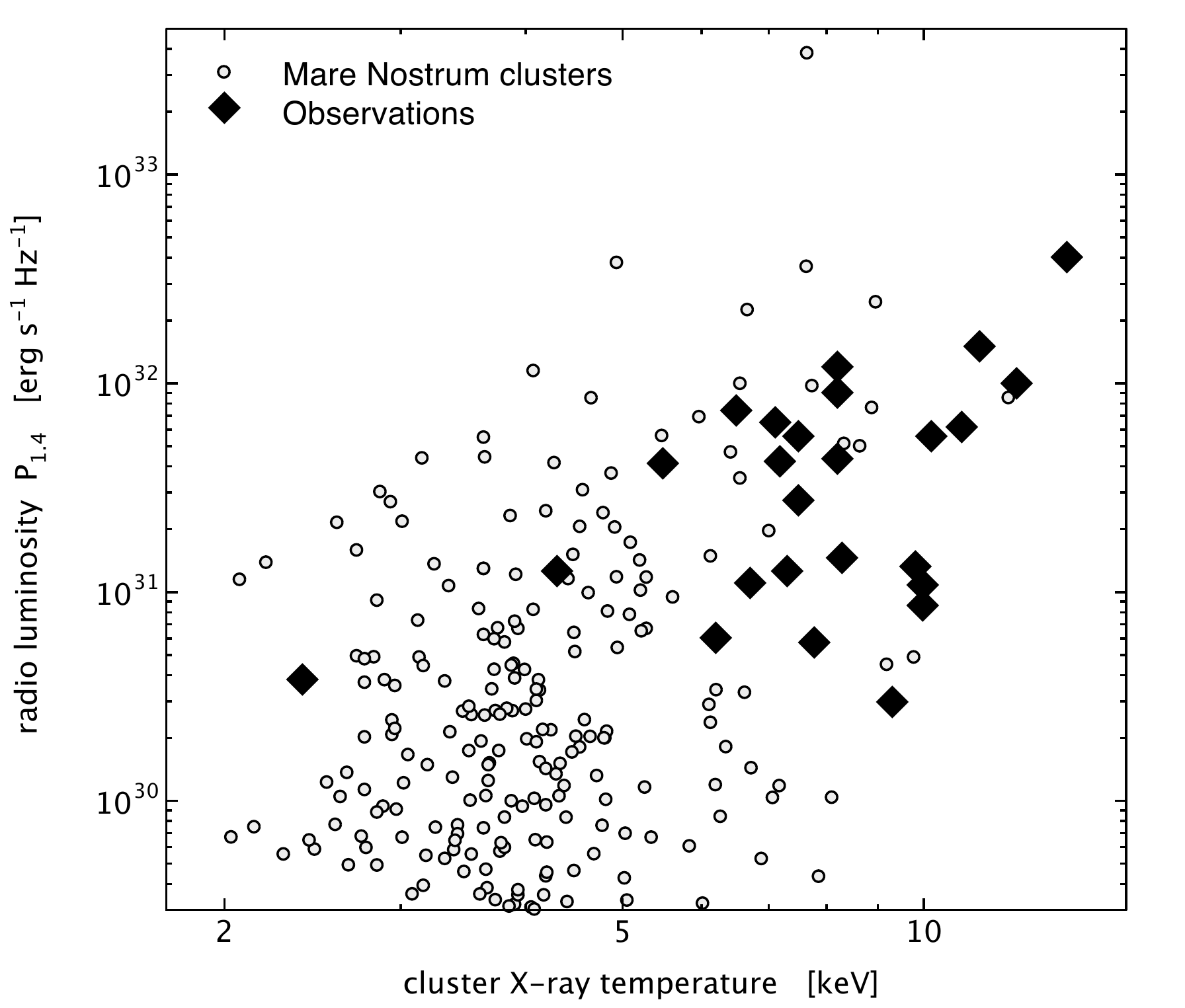}
	 \end{center}
	 \caption
	 { 
	 Bolometric X-ray temperature versus radio luminosity. For
	 comparison observational data are shown. The data include radio
	 relics and halos (\citet{govoni:01}, \citet{govoni:04},
	 \citet{feretti:97}, \citet{liang:00}, \citet{giovannini:99} ).
	 }
	 \label{fig-radio-xray}
   \end{figure}

   We also estimate the radio--X-ray correlation of the galaxy clusters
   in our simulation by computing the bolometric X-ray luminosity and
   the emission-weighted temperature. \Fig{fig-radio-xray} shows that
   only a fraction of the clusters is very luminous in radio, meaning
   that they show a radio luminosity above $10^{31} \, {\rm erg \,
   s^{-1} \, Hz^{-1} }$. This is also true for a significant number of
   the massive, \ie\ hot, clusters. However, the maximum radio
   luminosity scales strongly with the cluster temperature. This is
   indeed expected for radio relics the most massive clusters \citep[see
   e.g.][]{feretti:04b}. We plot for comparison the radio luminosity of
   several radio relics and halos. The \MNUniverse\ simulation clusters
   show a trend similar to the observed ones, however, the radio
   luminosity of our clusters is in average higher than that of the
   observed clusters. This difference may be caused by our assumptions
   for the parameters in the radio emission model, \ie\ the shock
   acceleration is too efficient in our model, or the magnetic fields
   are too strong. Alternatively, the X-ray temperature of the simulated
   clusters might be too low. In a forthcoming work we will use the
   comparison to observed radio halos and relics to constrain our model
   parameters. Here we wish to point out that our simple emission model
   taking only primary electrons into account nicely restores in
   combination with a cosmological simulation the fact that only a small
   fraction of clusters show luminous diffuse radio structures and in
   addition the trend in the radio luminosity--X-ray relation.

\subsection{Radio emission of accretion shocks}   
\label{sec-accretion-shock}

   \begin{figure*}
	 \begin{center}
	 \includegraphics[width=0.9\textwidth,angle=0]{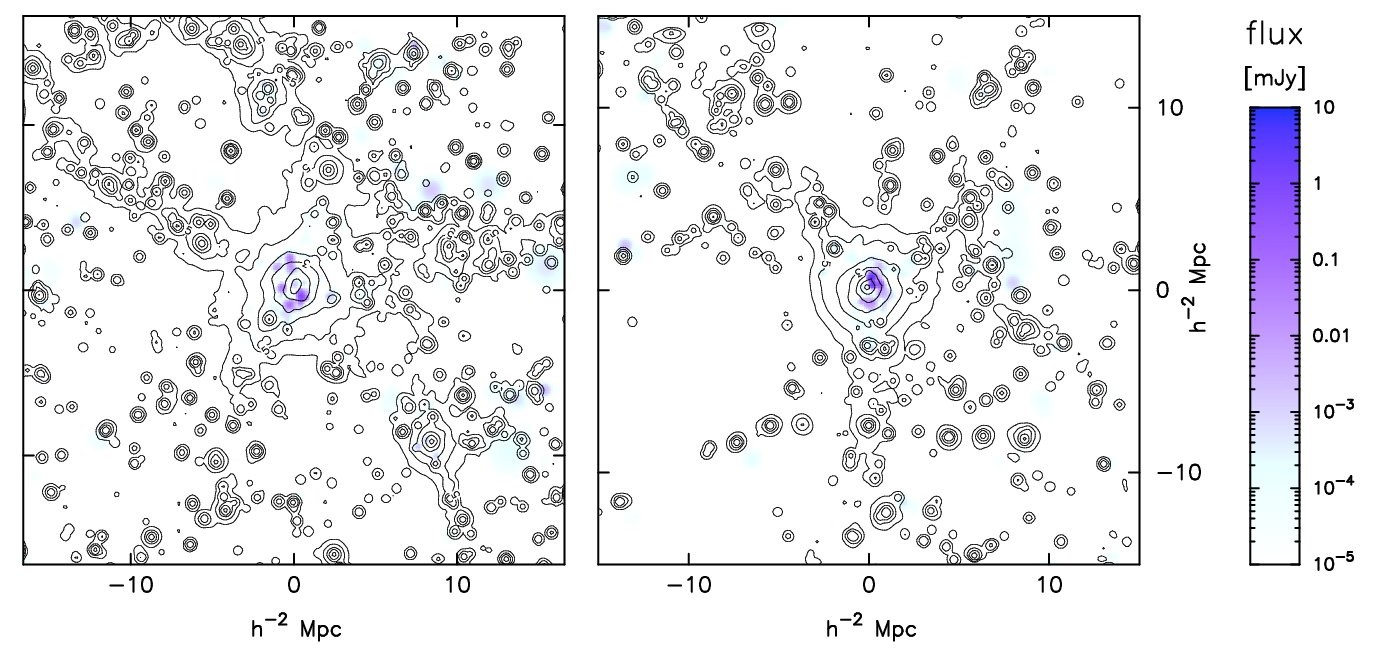}
	 \end{center}
	 \caption
	 { 
	 Mimic observations of the X-ray and the radio emission of the
	 periphery of the two most massive clusters in the \MNUniverse\
	 simulation.
	 }
	 \label{fig-radio-accretion}
   \end{figure*}

   Our method also allows to study the radio emission of accretion
   shocks. Here we show artificial radio maps for two of the most
   massive clusters in the simulation using a larger field-of-view, see
   \Fig{fig-radio-accretion}. We find only very little emission in the
   region of the accretion shocks, which are roughly located at four
   times the virial radius, \ie\ $ \sim 10 \, h^{-1} \, {\rm Mpc}$ for
   the two most massive clusters. The reason is that the temperature,
   density and magnetic field are simultaneously low in the cluster
   periphery. As a result the two clusters shown in
   \Fig{fig-radio-accretion} have a radio emission even below
   $10^{-4}\,{\rm mJy}$ (for a $10'' \times 10''$ beam). Hence, it may
   be difficult even with the sensitivity of upcoming radio telescopes
   to detect the accretion shock by their radio emission. 
   \\

\subsection{Magnetic field strength in radio relics}
\label{sec-relicB}

   Magnetic fields in the region of radio relics are difficult to
   determine observationally since the field strength and the number
   density of relativistic electrons are degenerated in the expression
   for the radio luminosity. Estimates for ICM magnetic fields range
   from 0.1 to $10 \, {\rm \mu G}$. We compute the radio
   luminosity-averaged magnetic field within two times the virial radius
   for the sample of clusters presented here in detail, see
   \Tab{tab-clusters}.  More precisely, we compute \begin{equation}
   \langle B\down \rangle = \frac{1}{P_{1.4}(<2 r_{\rm vir}) }
   \sum\nolimits_{r_i<2r_{\rm vir} } \, P_{1.4}(\vec{r}_i) B\down
   (\vec{r}_i) \nonumber .  \end{equation} We find that that for the
   clusters presented here the average magnetic field strength is in the
   range of 0.07 to $0.8 \, {\rm \mu G}$. Despite general
   over-luminosity of of the radio objects generated in our simulation
   the magnetic fields in the downstream region has been modelled with
   moderate field strength only. This implies that the abundance of
   radio relics in general can be explained with rather weak magnetic
   fields in the periphery of galaxy clusters.

\section{Discussion and Summary}
\label{sec-discussion}

   We have presented a novel method for estimating the radio emission of
   strong shocks that occur during the process of structure formation in
   the universe. Our approach is based on an estimate for the shock
   surface area and we compute the radio emission per surface element
   using the Mach number of the shock and the downstream plasma
   properties. The advantage of our method is that we can accurately
   determine the emission of the downstream regime without using an
   approximate, discretised electron spectrum. The disadvantage is
   clearly that we do not include any evolution of the shock front and
   the downstream medium. However, our main goal here is to compute the
   radio emission from all shock fronts generated during cosmic
   structure formation. Our main assumption is that radio emission is
   caused by primary electrons accelerated at the shock front. To obtain
   a model suitable for the application in the framework of a
   cosmological simulation we have to neglect the complexity of
   collisonless shocks, details of the electron acceleration mechanisms,
   the amplification of magnetic fields by upstream cosmic rays, and so
   forth. The model reflects rather an expectation for the average radio
   emission entailed by structure formation shocks.  
   \\

   Our results show that in a cosmological simulation one indeed finds
   textbook examples for large-scale, ring-like radio relics as observed
   in A3667 and A3376. Our example {\em A} shows two large half-shell
   shaped shock fronts and no radio emission in the centre of the
   merging cluster. However, cluster {\em A} is not very massive, hence
   the radio luminosity of the two half-shells is low compared to A3667.
   Another nice example is cluster {\em B} which shows a radio relic
   only one side of the cluster, similar to A115. Beside those
   spectacular relics in the periphery of galaxy clusters, we find that
   a lot of clusters show complex, radio-loud shock structures close to
   the cluster centre. This suggests that part of the central diffuse
   radio emission in galaxy clusters, \ie the radio halo phenomena, can
   be attributed to synchrotron emission of primary electrons.
   \\

   We have modelled the radio emission with conservative assumptions for
   the efficiency of shock acceleration, namely $\xi_e = 0.005$, and for
   the strength of magnetic fields in the downstream region, namely of
   the order of 0.07 to $0.8\,{\rm \mu G}$. Still, we find that the
   radio luminosity in the cluster region is on average significantly
   above observed values, see \Fig{fig-cum-synchro}. Even lower values
   for the acceleration efficiency or the magnetic field strength would
   suffice to explain the abundance of diffuse radio objects on the sky.
   In conclusion, our findings support the scenario in which radio
   relics are generated by primary electrons. 
   \\

   In summary, we have analysed one of the largest hydrodynamical
   simulations of cosmic structure formation with respect to shock
   fronts in clusters of galaxies. We have developed a method to
   identify the shock fronts, to estimate their Mach numbers and their
   orientation. We have applied the method described in \citet{hoeft:07}
   to determine the radio emission as a function of the shock surface
   area and the downstream plasma properties. Our analysis led to the
   following results:
   
   \begin{itemize}
   
   \item
   By evaluating the entropy and the velocity field we are able to
   identify strong shocks in the simulation and to determine their
   properties.
  
   \item
   Using conservative values for the efficiency of strong shocks to
   accelerate electrons and for strengths of magnetic fields, we are
   able to reproduce the number density and the luminosity of
   large-scale radio relics. Only a very few of the most massive
   clusters show such luminous, extended sources.

   \item
   Our model reproduces various spectacular sources of diffuse radio
   emission at the periphery of galaxy clusters. Other clusters show
   emission with complex morphology close to the cluster centre. These
   sources may be classified as part of a central radio halo, following the
   observational distinction between relics and halos.

   \item
   Our results reproduce the strong correlation between radio luminosity
   and cluster temperature. The highest radio luminosities occur in
   dense and hot environments such as massive clusters. In addition
   we find that a large number of galaxy clusters show only little
   diffuse radio emission.

   \item
   We find that the abundance of radio relics can be explained with
   efficiency of diffusive shock acceleration for electrons lower than
   $\xi_e = 0.005$ and a strength of magnetic fields in the relic region
   lower than 0.07 to $0.8 \, { \rm \mu G}$.

   \item 
   All of the luminous radio relics belong to internal shock fronts. The
   accretion shocks are located at larger distances from the cluster
   centre. Since density and temperature are low at this location, their
   luminosity is too small to reach a similar flux as internal shock
   fronts.

   \end{itemize}

\noindent {\sc Acknowledgment }

   MH acknowledges DLR funding under the grant 50 OX 0201. MB
   acknowledges the support by the DFG grant BR 2026/3. The \MNUniverse\
   simulations have been done at the Barcelona Supercomputing Center
   (Spain) and analysed at NIC J\"ulich (Germany). GY thanks MEC (Spain)
   for financial support under project numbers FPA2006-01105 and
   AYA2006-15492-C03. Our collaboration has been supported by the
   European Science Foundation (ESF) for the activity entitled
   `Computational Astrophysics and Cosmology' ({\sc AstroSim}).

\newcommand{\aap  }{A\&A}
\newcommand{\araa }{ARA\&A}
\newcommand{\apj  }{ApJ}
\newcommand{\apjs }{ApJS}
\newcommand{\apjl }{ApJL}
\newcommand{\apss }{ApSS}
\newcommand{\aapr }{A\&A~Rev.}
\newcommand{\aj   }{AJ}
\newcommand{\jgr  }{JGR}
\newcommand{\mnras}{MNRAS}
\newcommand{\nat   }{Nature}
\newcommand{\physrep  }{PhysRep}

\bibliography{radio}
\bibliographystyle{apj}

\end{document}